\def\year{2018}\relax
\documentclass[letterpaper]{article} 
\usepackage{aaai18}  
\usepackage{times}  
\usepackage{helvet}  
\usepackage{courier}  
\usepackage{url}  
\usepackage{graphicx}  
\usepackage[english]{babel}
\usepackage[utf8x]{inputenc}
\usepackage[T1]{fontenc}

\usepackage{amsmath,amssymb,amsfonts,bbm}
\usepackage{dsfont} 
\usepackage[svgnames]{xcolor}
\usepackage{graphicx}
\usepackage[colorinlistoftodos]{todonotes}
\usepackage{url}
\usepackage{soul}

\usepackage{algorithm}
\usepackage{algorithmic}
\usepackage{array}
\usepackage{indentfirst}

\usepackage{booktabs} 
\usepackage{etoc} 
\usepackage{multirow}
\usepackage{makecell}
\usepackage{enumitem}

\usepackage{verbatim}
\usepackage[export]{adjustbox}
\usepackage{subfig}
\usepackage{epstopdf}
\DeclareGraphicsExtensions{.png,.pdf,.eps,.jpg,.tif,.tiff,.ps} 
\graphicspath{{img//}{img/yifei-work//}{img/wordclouds//}}

\newcommand{\citet}[1]{\citeauthor{#1}~\shortcite{#1}}
\newcommand{\citep}{\cite}

\usepackage{soul}
\usepackage[colorinlistoftodos]{todonotes}

\definecolor{navy}{rgb}{0.1, 0.1, 0.8}
\definecolor{gray}{rgb}{0.6, 0.6, 0.6}
\definecolor{myblue}{rgb}{.8, .8, 1}
\definecolor{olive}{rgb}{0.1, 0.5, 0.1}

\usepackage{xspace}
\newcommand{\debate}{{\sc \#DebateNight}\xspace}
\newcommand{\Protected}{\texttt{Protected}\xspace}
\newcommand{\Human}{\texttt{Human}\xspace}
\newcommand{\Bot}{\texttt{Bot}\xspace}
\newcommand{\Suspended}{\texttt{Suspended}\xspace}

\begin{document}
\etocdepthtag.toc{mtchapter}
%
\title{\debate: The Role and Influence of Socialbots on Twitter\\During the 1st 2016 U.S. Presidential Debate}
\author{
	Marian-Andrei Rizoiu\textsuperscript{1}\textsuperscript{2} 
	\and Timothy Graham\textsuperscript{1} 
	\and Rui Zhang\textsuperscript{1}\textsuperscript{2}  \\
	{\bf \Large \and Yifei Zhang\textsuperscript{1}\textsuperscript{2} \and Robert Ackland\textsuperscript{1} \and Lexing Xie\textsuperscript{1}\textsuperscript{2}} \\
	\textsuperscript{1}The Australian National University,
	\textsuperscript{2}Data61 CSIRO, 
	Canberra, Australia.\\
}

\maketitle

\begin{abstract}
%
Serious concerns have been raised about the role of `socialbots' in manipulating public opinion and influencing the outcome of elections by retweeting partisan content to increase its reach.
Here we analyze the role and influence of socialbots on Twitter by determining how they contribute to retweet diffusions.
We collect a large dataset of tweets during the 1st U.S. presidential debate in 2016 and we analyze its 1.5 million users from three perspectives: user influence, political behavior (partisanship and engagement) and botness.
First, we define a measure of user influence based on the user's active contributions to information diffusions, i.e. their tweets and retweets. 
Given that Twitter does not expose the retweet structure -- it associates all retweets with the original tweet --
we model the latent diffusion structure using only tweet time and user features, and we implement a scalable novel approach to estimate influence over all possible unfoldings.
Next, we use partisan hashtag analysis to quantify user political polarization and engagement.
Finally, we use the BotOrNot API to measure user \emph{botness} (the likelihood of being a bot).
We build a two-dimensional ``polarization map'' that allows for a nuanced analysis of the interplay between botness, partisanship and influence.
We find that not only are socialbots more active on Twitter -- starting more retweet cascades and retweeting more -- but they are 2.5 times more influential than humans, and more politically engaged. 
Moreover, pro-Republican bots are both more influential and more politically engaged than their pro-Democrat counterparts.
However we caution against blanket statements that software designed to appear human dominates politics-related activity on Twitter.
Firstly, it is known that accounts controlled by teams of humans (e.g. organizational accounts) are often identified as bots.
Secondly, we find that many highly influential Twitter users are in fact pro-Democrat and that most pro-Republican users are mid-influential and likely to be human (low botness).
\end{abstract}


\section{Introduction}


Socialbots are broadly defined as ``software processes that are programmed to appear to be human-generated within the context of social networking sites such as Facebook and Twitter'' \cite[p.2]{GehlBakardjieva2016}.
They have recently attracted much attention and controversy, with concerns that they infiltrated political discourse during the 2016 U.S. presidential election and manipulated public opinion at scale. 
Concerns were heightened with the discovery that an influential conservative commentator (\textit{@Jenn\_Abrams}, 70,000 followers) and a user claiming to belong to the Tennessee Republican Party (\textit{@TEN\_GOP}, 136,000 followers) 
-- both retweeted by high-profile political figures and celebrities -- were in fact Russian-controlled bots operated by the Internet Research Agency in St. Petersburg~\cite{abramsDailyBeast,tenGopArticle}.


There are several challenges that arise when conducting large-scale empirical analysis of political influence of bots on Twitter. 
The first challenge concerns estimating user influence from retweet diffusions, where the retweet relations are unobserved -- the Twitter API assigns every retweet to the original tweet in the diffusion.
Current state-of-the-art influence estimation methods such as ConTinEst~\cite{Du2013} operate on a static snapshot of the diffusion graph, which needs to be inferred from retweet diffusions using approaches like NetRate~\cite{Gomez-Rodriguez2011}.
This workflow suffers from two major drawbacks: 
first, the algorithms for uncovering the diffusion graph do not scale to millions of users like in our application;
second, operating on the diffusion graph estimates the ``potential of being influential'', but it loses information about user activity -- e.g. a less well connected user can still be influential if they tweet a lot.
The question is \textbf{how to estimate at scale the influence of millions of users from diffusion in which the retweet relation is not observed?}
%
%
The second challenge lies in determining at scale whether a user is a bot and also her political behavior, as manually labeling millions of users is infeasible.
The question is therefore \textbf{how to leverage automated bot detection approaches 
to measure the botness of millions of users?
and how to analyze political behavior (partisanship and engagement) at scale?}

This paper addresses the above challenges using a large dataset (hereafter referred to as \debate) of 6.5 million tweets authored by 1.5 million users that was collected on 26 September 2016 during the 1st U.S. presidential debate.

To address the first challenge, we introduce, evaluate, and apply a novel algorithm to estimate user influence based on retweet diffusions.
We model the latent diffusion structure using only time and user features by introducing the \emph{diffusion scenario} -- a possible unfolding of a diffusion -- and its likelihood.
We implement a scalable algorithm to estimate user influence over all possible diffusion scenarios associated with a diffusion.
We demonstrate that our algorithm can accurately recover the ground truth on a synthetic dataset.
We also show that, unlike simpler alternative measures--such as the number of followers, or the mean size of initiated cascades--our influence measure $\varphi$ assigns high scores to both highly-connected users who never start diffusions and to active retweeters with little followership.

We address the second challenge by proposing three new measures (political polarization $\mathcal{P}$, political engagement $\mathcal{E}$ and botness $\zeta$) and by computing them for each user in \debate.
We manually compile a list of partisan hashtags and we estimate political engagement based on the tendency to use these hashtags and political polarization based on whether pro-Democrat or pro-Republican hashtags were predominantly used.
We use the BotOrNot API to evaluate botness and to construct four reference populations -- \Human, \Protected, \Suspended and \Bot.
We build a two-dimensional visualization -- the \emph{polarization map} -- that enables a nuanced analysis of the interplay between botness, partisanship and influence.
We make several new and important findings: 
(1) bots are more likely to be pro-Republican; 
(2) bots are more engaged than humans, and pro-Republican bots are more engaged than pro-Democrat bots; 
(3) the average pro-Republican bot is twice as influential as the average pro-Democrat bot; 
(4) very highly influential users are more likely to be pro-Democrat; and 
(5) highly influential bots are mostly pro-Republican.


%
%
%

\textbf{The main contributions of this work include:}
\begin{itemize}
	\item We introduce a \textbf{scalable algorithm to estimate user influence} over all possible unfoldings of retweet diffusions where the cascade structure is not observed;
	the code is publicly accessible in a Github repository\footnote{Code and partisan hashtag list is publicly available at \url{https://github.com/computationalmedia/cascade-influence}};

	\item We develop two \textbf{new measures of political polarization and engagement} based on usage of partisan hashtags;
	the list of partisan hashtags is also available in the repository;
	
	\item We measure the \textbf{botness of a very large population of users} engaged in Twitter activity relating to an important political event -- the 2016 U.S presidential debates;
		
	\item We propose the \textbf{polarization map} -- a novel visualization of political polarization as a function of user influence and botness -- and we use it to gain insights into the influence of bots on the information landscape around the U.S. presidential election.
\end{itemize}


\section{Related work}

We structure the discussion of previous work into two categories:
related work on the estimation of user influence and
work concerning bot presence and behavior on Twitter.

\textbf{Estimating user influence on Twitter.}
Aggregate measures such as the follower count, the number of retweets and the number of mentions have been shown to be indicative of user influence on Twitter~\cite{Cha2010,kwak2010twitter}.
More sophisticated estimates of user influence use eigenvector centrality to account for the connectivity of followers or retweeters; 
for example, TwitteRank~\cite{twitterrank} extends PageRank~\cite{pagerank} by taking into account topical similarity between users and network structure.
Other extensions like Temporal PageRank~\cite{rozenshtein2016temporal} explicitly incorporate time into ranking to account for a time-evolving network.
However, one limitation of PageRank-based methods is that they require a complete mapping of the social networks.
More fundamentally, network centrality has the drawback of evaluating only the \textit{potential} of a user to be influential in spreading ideas or content, and it does not account for the actions of the user (e.g. tweeting on a particular subject).
Our influence estimation approach proposed in Sec.~\ref{sec:user-influence} is built starting from the user Twitter activity and it does not require knowledge of the social network.

Recent work~\cite{ICWSM1613006,Chikhaoui:2017:DCA:3127339.3070658} has focused on estimating user influence as the contribution to information diffusion. 
For example, ConTinEst~\cite{Du2013} requires a complete diffusion graph and employs a random sampling algorithm to approximate user influence with scalable complexity.
However, constructing the complete diffusion graph might prove problematic, as current state-of-the-art methods for uncovering the diffusion structure (e.g. \cite{Gomez-Rodriguez2011,simma2012modeling,cho2013latent,li2013dyadic,linderman2014discovering}) do not scale to the number of users in our dataset. 
This is because these methods assume that a large number of cascades occur in a rather small social neighborhood, whereas in \debate cascades occur during a short period of time in a very large population of users.
Our proposed algorithm estimates influence directly from retweet cascades, without the need to reconstruct the retweet graph, and it scales cubically with the number of users.


\begin{figure*}[tbp]
	\newcommand\mywidth{0.19}
	\centering
	\subfloat[] {
		\includegraphics[width=0.14\textwidth,valign=c]{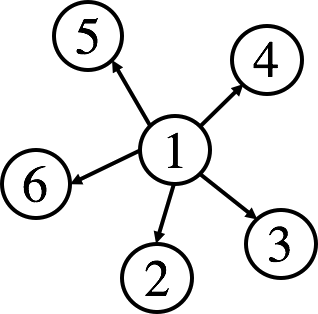}
		\vphantom{\includegraphics[height=\mywidth\textheight,valign=c]{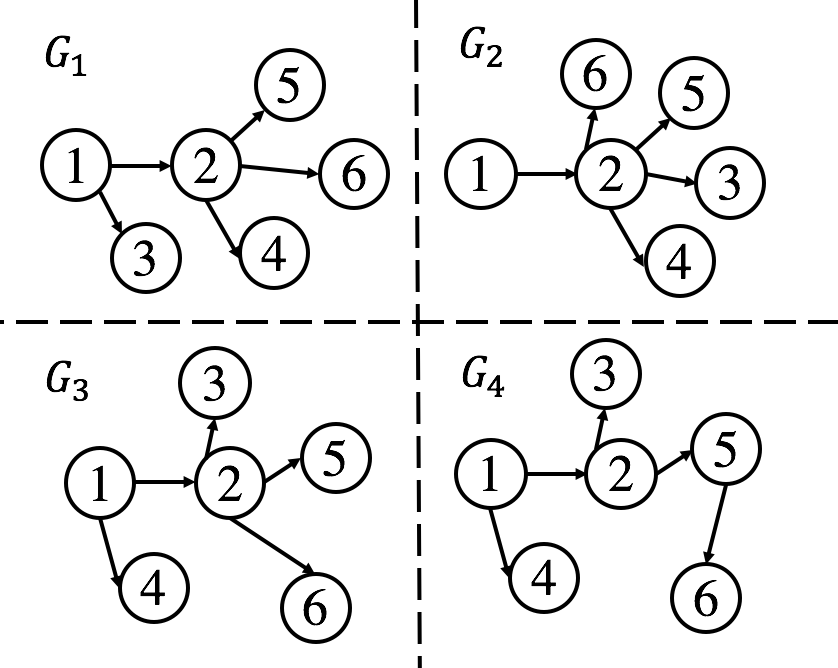}}
		\label{fig:side:a}
	}
	\hspace{0.15cm}
	\subfloat[] {
		\includegraphics[height=\mywidth\textheight,valign=c]{somepossicas}
		\label{fig:side:b}
	}
	\hspace{0.15cm}
	\subfloat[] {
		\includegraphics[height=\mywidth\textheight,valign=c]{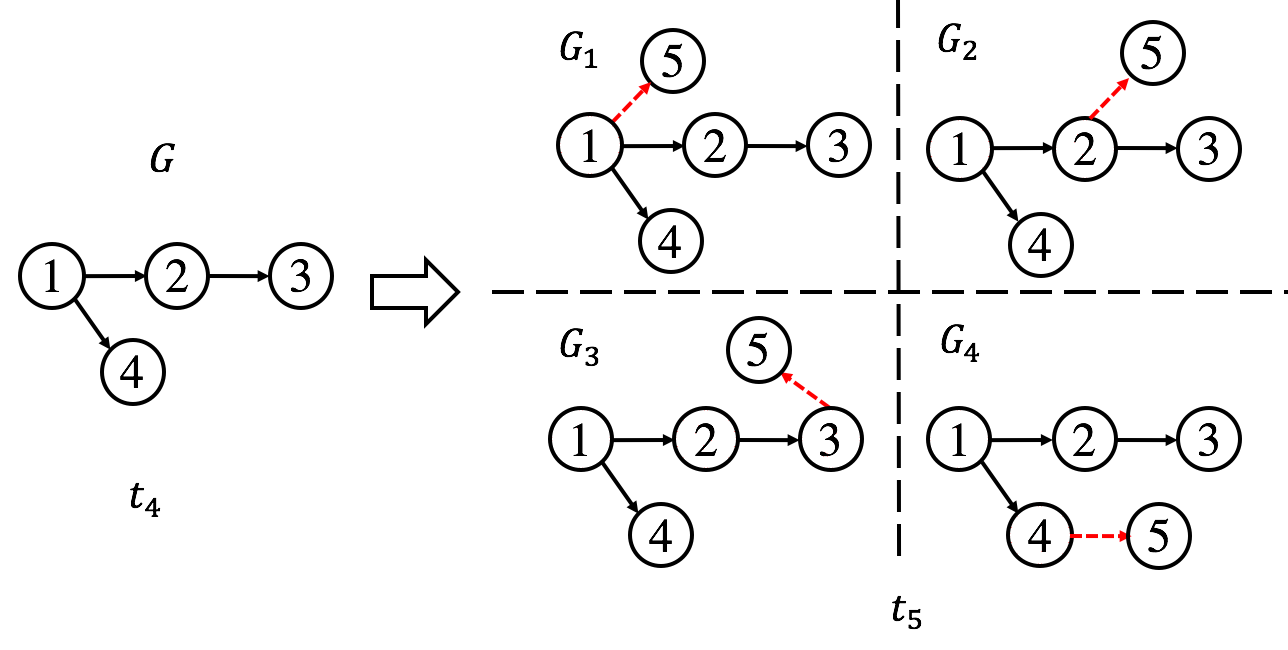}
		\label{fig:add-one-edge}
	}
	\caption{ 
		Modeling latent diffusions.
		\textbf{(a)} The schema of a retweet cascade as provided by the Twitter API, in which all retweets are attributed to the original tweet.
		\textbf{(b)} Four diffusion scenarios (out of 120 possible scenarios), associated with the retweet cascade in (a).
		\textbf{(c)} Intuition of the independent conditional model.
		A new node $v_5$ appears conditioned on one diffusion scenario $G$.
		Four new diffusion scenarios are generated as $v_5$ can attach to any of the existing nodes.
	}
	\label{fig:holdout-ll}
	\vspace{-1mm}
\end{figure*}

\textbf{Bot presence and behavior on Twitter.}
\label{previousworkbots}
The `BotOrNot' Twitter bot detection API uses a Random Forest supervised machine learning classifier to calculate the likelihood of a given Twitter user being a bot, based on more than 1,000 features extracted from meta-data, patterns of activity, and tweet content (grouped into six main classes: user-based; friends; network; temporal; content and language; and sentiment) 
\cite{davisetal.16,varol.17}\footnote{See: \url{https://botometer.iuni.iu.edu/\#!/}}.
The bot scores are in the range $[0,1]$, where 0 (1) means the user is very unlikely (likely) to be a bot. 
BotOrNot was used to examine how socialbots affected political discussions on Twitter during the 2016 U.S. presidential election~\cite{FM7090}.
They found that bots accounted for approximately 15 \% (400,000 accounts) of the Twitter population involved in election-related activity, and authored about 3.8 million (19 \%) tweets. 
However, \citet{FM7090} sampled the most active accounts, which could bias upwards their estimate of the presence of bots as activity volume is one of the features that is used by BotOrNot. 
They found that bots were just as effective as humans at \textit{attracting retweets} from humans. 
\citet{Woolley.2017} used BotOrNot to test 157,504 users randomly sampled from 1,798,127 Twitter users participating in election-related activity 
and found that over 10\% were bots. 
Here we use BotOrNot to classify \textit{all} 1.5 million users in our dataset to obtain a less biased approximation of their numbers and impact.

Previous work has studied the political partisanship of Twitter bots.
\citet{Kollanyi.2016.presidentialdebate} analyzed candidate-oriented hashtag use during the 1st U.S. presidential debate and found that highly automated accounts (self-identified bots and/or accounts that post at least 50 times a day) were disproportionately pro-Trump.
\citet{FM7090} also studied political partisanship by identifying five pro-Trump and four pro-Clinton hashtags and assigning users to a particular political faction.
The results suggested that both humans and bots were more pro-Trump in terms of hashtag partisanship.
However, the above findings are limited to a comparison between humans and bots of frequency counts of tweets authored and retweets received, and they provide no insight into the importance of users in retweet diffusions. 
We overcome this limitation by modeling the latent structure of retweet diffusions and computing user influence over all possible scenarios.

\section{Estimating influence in retweet cascades}
\label{sec:user-influence}



An {\em information cascade} $V$ of size $n$ is defined as a series of messages $v_i$ sent by user $u_i$ at time $t_i$, i.e. $V=\{v_i=(u_i, t_i)\}_{i=1:n}$.
Here $v_1 = (u_1, t_1)$ is the initial message, and $v_1, \ldots, v_n$ with $t_1<\ldots<t_n$ are subsequent reposts or relays of the initial message.
In the context of Twitter, the initial message is an original tweet and the subsequent messages are retweets of that original tweet (which by definition, are also tweets).
A latent retweet diffusion graph $G=(V,E)$ has the set of tweets as its vertexes $V$, and additional edges $E=\{(v_i, v_j)\}$ that represent that the $j^{th}$ tweet is a retweet of the $i^{th}$ tweet, and respects the temporal precedence $t_i<t_j$. 
Web data sources such as the Twitter API provide cascades, but not the
diffusion edges. 
Such missing data makes it challenging to measure a given user's contribution to the diffusion process.



\subsection{Modeling latent diffusions}
\label{subsec:model-latent-diffusions}

\noindent\textbf{Diffusion scenarios.}
We focus on tree-structured diffusion graphs, i.e. each node $v_j$ has only one incoming link $(v_i, v_j)$, $i<j$. 
Denote the set of trees that are consistent with the temporal order in cascade $C$ as ${\cal G}$, we call each diffusion tree a \emph{diffusion scenario} $G \in {\cal G}$.
Fig.~\ref{fig:side:a} contains a cascade visualized as a star graph, attributing subsequent tweets to the first tweet at $t_1$.
Fig.~\ref{fig:side:b} shows four example diffusion scenarios consistent with this cascade.
The main challenge here is to estimate the influence of each user in the cascade, taking into account all possible diffusion trees.

\noindent\textbf{Probability of retweeting.}
For each tweet $v_j$, we model the probability of it being a direct descendant of each previous tweet in the same cascade as a weighted softmax function, defined by two main factors:
firstly, users retweet \emph{fresh} content~\cite{Wu2007}.
We assume that the probability of retweeting decays exponentially with the time difference $t_j - t_i$;
secondly, users prefer to retweet locally influential users, known as preferential attachment~\cite{Barabasi2005,Rizoiu2017}.
We measure the local influence $m_i$ of a user $u_i$ using her number of followers~\cite{kwak2010twitter,Cha2010}.
We quantify the probability that $v_j$ is a direct retweet of $v_i$ as:
\begin{equation} \label{eq:prob-edge-mt}
	p_{ij} = \frac{m_i e^{-r({t_j-t_i})}}{\sum_{k=1}^{j-1} m_k e^{-r({t_j-t_k})}}
\end{equation}
where 
$r$ is a hyper-parameter controlling the temporal decay. 
It is set to $r = 6.8 \times 10^{-4}$, tuned using linear search on a sample of 20 real retweet cascades (details in the supplement~\cite[annex~D]{supplemental}).

\subsection{Tweet influence in a retweet cascade}
\label{subsec:user-influence-mt}


We additionally assume retweets follow {\em independent conditional diffusions} within a cascade. 
This is to say that conditioned on an existing partial cascade of $j-1$ retweets denoted as $V^{(j-1)}=\{v_k\}_{k=1}^{j-1}$ whose underlying diffusion scenario is $G^{(j-1)}$, the $j^{th}$ retweet is attributed to any of the $k=1,\ldots,j-1$ prior tweets according to Eq.~\ref{eq:prob-edge-mt}, and is independent of the diffusion scenario $G^{(j-1)}$. 
For example, the $5^{th}$ tweet in the cascade will incur four valid diffusion trees for each of the diffusion scenarios for 4 tweets -- this is illustrated in Fig.~\ref{fig:add-one-edge}. 
This simplifying assumption is reasonable, as it indicates that each user $j$ makes up his/her own mind about whom to retweet, and that the history of retweets is available to user $j$ (as is true in the current user interface of Twitter).
It is easy to see that under this model, the total number of valid diffusion trees for a 5-tweet cascade is $1\cdot 2\cdot 3\cdot 4=24$, and that for a cascade with $n$ tweets is $(n-1)!$.

The goal for influence estimation for each cascade is to compute the contribution $\varphi(v_i)$ of each tweet $v_i$ {\em averaging} over all independent conditional diffusion trees consistent with cascade $V$ and with edge probabilities prescribed by Eq.~\ref{eq:prob-edge-mt}. 
Enumerating all valid trees and averaging is clearly computationally intractable, but the illustration in Fig.~\ref{fig:add-one-edge} lends itself to a recursive algorithm. 

\textbf{Tractable tweet influence computation}
We introduce the \emph{pair-wise influence score} $m_{ij}$ which measures the influence of $v_i$ over $v_j$.
$v_i$ can influence $v_j$ both directly when $v_j$ is a retweet of $v_i$, and indirectly when a path exists from $v_i$ to $v_j$ in the underlying diffusion scenario.
Let $v_k$ be a tweet on the path from $v_i$ to $v_j$ ($i < k < j$) so that $v_j$ is a direct retweet of $v_k$.
$m_{ik}$ can be computed at the $k^{th}$ recursion step and it measures the influence of $v_i$ over $v_k$ over all possible paths starting with $v_i$ and ending with $v_k$.
Given the above independent diffusions assumption, the $m_{ij}$ can be computed using $m_{ik}$ to which we add the edge $(v_k, v_j)$.
User $u_j$ can chose to retweet any of the previous tweets with probability $p_{kj}, k < j$, therefore we further weight the contribution through $v_k$ using $p_{ij}$.
We consider that a tweet has a unit influence over itself ($m_{ii} = 1$).
Finally, we obtain that:
\begin{equation} \label{eq:Mij-mt}
m_{ij} =
\left\{
\begin{array}{ll}
	\sum^{j-1}_{k=i} m_{ik}p_{kj}^2 &,i < j \\
	1 & ,i = j \\
	0 & ,i > j.
\end{array}
\right.
\end{equation}

Naturally, $\varphi(v_i)$ the total influence of node $v_i$ is the sum of $m_{ij}, j > i$ the pair-wise influence score of $v_i$ over all subsequent nodes $v_j$. 
The recursive algorithm has three steps. 
\begin{enumerate}
	\item {\bf Initialization.} $m_{ij}=0$ for $i, j=1,\ldots,n, j \neq i$, and $m_{ii} = 1$ for $i=1,\ldots,n$;
	\item {\bf Recursion.} For $j=2, \ldots, n$;
	\begin{enumerate}
		\item For $k=1, \ldots, j-1$, compute $p_{kj}$ using Eq.~\eqref{eq:prob-edge-mt};
		\item For $i=1, \ldots, j-1$, $m_{ij} = \sum_{k=i}^{j-1} m_{ik}p_{kj}^2$;
	\end{enumerate}
	\item {\bf Termination.} Output $\varphi(v_i)= \sum_{k=i+1}^n m_{ik}$, for $i=1,\ldots,n$. 
\end{enumerate}

We exemplify this algorithm on a 3-tweet toy example. 
Consider the cascade $\{v_1, v_2, v_3\}$.
When the first tweet $v_1$ arrives, we have $m_{11} = 1$ by definition (see Eq.~\eqref{eq:Mij-mt}).
%
After the arrival of the second tweet, which must be retweeting the first, we have 
$m_{12}= m_{11} p^2_{12}=1$, and $m_{22} = 1$ by definition.
The third tweet can be a retweet of the first or the second, therefore we obtain: 
\begin{align*}
	m_{13} =& m_{11} p^2_{13} + m_{12} p^2_{23} \; ;\\
	m_{23} =& m_{22} p^2_{23} \; ;\\
	m_{33} =& 1 \;.
\end{align*}
The second term of $m_{13}$ accounts for the indirect influence of $v_1$ over $v_3$ through $v_2$.
This is the final step for a 3-node cascade. 

The computational complexity of this algorithm is $O(n^3)$.
There are $n$ recursion steps, and calculating $p_{ij}$ at sub-step (a) needs $O(n)$ units of computation, and sub-step (b) takes $O(n^2)$ steps.
In real cascades containing 1000 tweets, the above algorithm finishes in 34 seconds on a PC.
For more details and examples, see the online supplement~\cite[annex~B]{supplemental}.

\subsection{Computing influence of a user}
\label{subsec:user-influence-measure}

Given $\mathcal{T}(u)$ -- the set of tweets authored by user $u$ --, we define the user influence of $u$ as the mean tweet influence of tweets $v \in \mathcal{T}(u)$:
\begin{equation} \label{eq:user-infl-casin}
	\varphi(u) = \frac{\sum_{v \in \mathcal{T}(u)} \varphi(v)}{|\mathcal{T}(u)|}, \mathcal{T}(u) = \{ v | u_v = u\}
\end{equation}
To account for the skewed distribution of user influence, we mostly use the normalization --  percentiles with a value of 1 for the most influential user our dataset and 0 for the least influential -- denoted $\varphi(u) \%$ .


\section{Dataset and measures of political behavior}
\label{sec:four-measures}

In this section, we first describe the \debate dataset that we collected during the 1st U.S. presidential debate.
Next, we introduce three measures for analyzing the political behavior of users who were active on Twitter during the debate.
In Sec.~\ref{subsec:political-polarization-measures}, we introduce \emph{political polarization} $\mathcal{P}$ and \emph{political engagement} $\mathcal{E}$.
In Sec.~\ref{subsec:bot-detection} we introduce the \emph{botness score} $\zeta$ and we describe how we construct the reference bot and human populations.


\textbf{The \debate dataset}
contains Twitter discussions that occurred during the 1st 2016 U.S presidential debate between Hillary Clinton and Donald Trump.
Using the Twitter Firehose API\footnote{Via the Uberlink Twitter Analytics Service.}, we collected all the tweets (including retweets) that were authored during the two hour period from 8.45pm to 10.45pm EDT, on 26 September 2016, and which contain at least one of the hashtags: \texttt{\#DebateNight}, \texttt{\#Debates2016}, \texttt{\#election2016}, \texttt{\#HillaryClinton}, \texttt{\#Debates}, \texttt{\#Hillary2016}, \texttt{\#DonaldTrump} and \texttt{\#Trump2016}.
The time range includes the 90 minutes of the presidential debate, as well as 15 minutes before and 15 minutes after the debate.
The resulting dataset contains 6,498,818 tweets, emitted by 1,451,388 twitter users.
For each user, the Twitter API provides aggregate information such as the number of followers, the total number (over the lifetime of the user) of emitted tweets, authored retweets, and favorites.
For individual tweets, the API provides the timestamp and, if it is a retweet, the original tweet that started the retweet cascade.
The \debate dataset contains 200,191 retweet diffusions of size 3 and larger.

\subsection{Political polarization $\mathcal{P}$ and engagement $\mathcal{E}$}
\label{subsec:political-polarization-measures}


\begin{figure}[tbp]
	\centering
	\newcommand\mywidth{0.45}
	\includegraphics[width=\mywidth\textwidth]{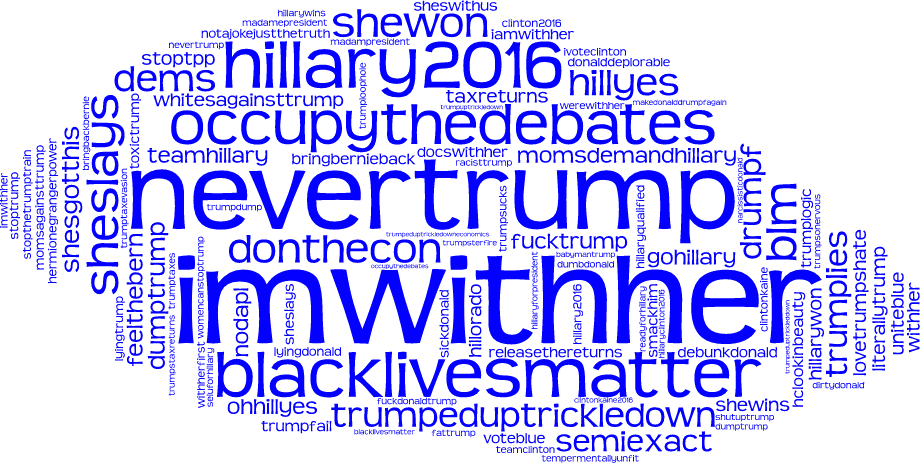} 
	\vspace{0.1cm}\\
	\includegraphics[width=\mywidth\textwidth]{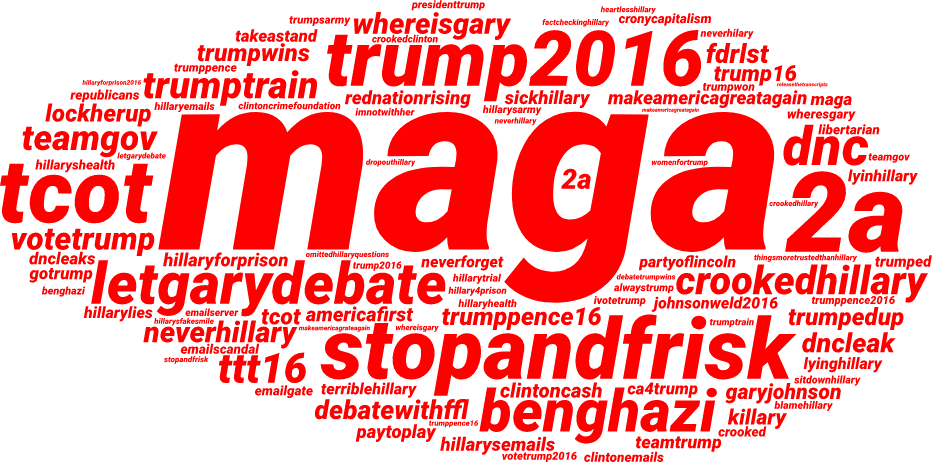}
	\caption{
		Wordclouds of partisan hashtags in \debate: Democrat \textbf{(top)} and Republican \textbf{(bottom)}.
		Hashtags sizes are scaled by their frequency.
	}
	\label{fig:wordclouds}
\end{figure}


\textbf{Protocol.}
Content analysis~\cite{kimkuljis2010} was used to code the 1000 most frequently occurring hashtags according to their political polarity. More specifically, we used 
Directed Content Analysis~\cite{hsieh-shannon-2005} to contextually analyse hashtags and code them according to their political polarity (or not, denoted as `neutral' and subsequently excluded from analysis). 
This approach has been used in previous work to study hashtags on Twitter in a manner that is valid, reliable and replicable~\cite{small-2011}. 
There were two previous studies of Twitter activity during the 2016 U.S. presidential election that informed the development of our coding schema. 
Firstly, \citet{FM7090} devised a binary classification scheme that attributed political partisanship to a small set of key hashtags as either `Trump-supporting' (\#donaldtrump, \#trump2016, \#neverhillary, \#trumppence16, \#trump) or `Clinton-supporting' (\#hillaryclinton, \#imwithher, \#nevertrump, \#hillary). 
Secondly, in studying Twitter activity during the 1st U.S. presidential debate, \citet{Kollanyi.2016.presidentialdebate} developed a coding schema that categorized tweets into seven categories based on the hashtags that occurred within the tweet. 
However, the authors found that three `exclusive' categories (`Pro-Trump', `Pro-Clinton', and `Neutral') accounted for the majority (88.5\%) of observations. 

Given the findings of previous research, we developed a code book with three categories: `Pro-Trump', `Pro-Clinton', and `Neutral'. 
To ensure that hashtags were analyzed within context, our content analysis methodology focussed on three units of analysis (following the approach developed by~\citet{small-2011}). 
The first is hashtags, comprised of a set of the 1000 most frequently occurring hashtags over all tweets in our dataset. 
The second unit of analysis was individual tweets that contained these hashtags. 
In order to gain a more nuanced and `situated' interpretation of hashtag usage, for each hashtag we referred to a small random sample of tweets in our dataset that contained each given hashtag. 
In some instances the polarity (or neutrality) was clear and/or already determined from previous studies, which helped to speed up the analysis of tweets. 
The third unit of analysis was user profiles, which we referred to in situations where the polarity or neutrality of a given hashtag was unclear from the context of tweet analysis. 
For example, \#partyoflincoln was used by both Republican and Democrat Twitter users, but an analysis of both tweets and user profiles indicated that this hashtag was \textit{predominantly} used by Pro-Trump supporters to positively align the Republican Party with the renowned historical figure of President Abraham Lincoln, who was a Republican. 
The content analysis resulted in a subset of 93 pro-Democrat and 86 pro-Republican hashtags (see the wordcloud visualization in Fig.~\ref{fig:wordclouds}), whilst the remaining `neutral' hashtags were subsequently excluded from further analysis.
The resulting partisan hashtag list contains hashtags indicating either strong support for a candidate (e.g., \texttt{\#imwithher} for Clinton and \texttt{\#trump2016} for Trump), or opposition and/or antagonism (e.g., \texttt{\#nevertrump} and \texttt{\#crookedhillary}).
The complete list of partisan hashtags is publicly available in the Github repository.


\textbf{Two measures of political behavior.}
We identify 65,031 tweets in \debate that contain at least one partisan hashtag (i.e., one of hashtags in the reference set of partisan hashtags constructed earlier).
1,917 tweets contain partisan hashtags with both polarities: these are mostly negative tweets towards both candidates (e.g., ``Let's Get READY TO RUMBLE AND TELL LIES. \#nevertrump \#neverhillary \#Obama'') or hashtag spam.
We count the number of occurrences of partisan hashtags for each user, and we detect a set of 46,906 politically engaged users that have used at least one partisan hashtag.
Each politically engaged user $u_i$ has two counts: $dem_i$ the number of Democrat hashtags that $u_i$ used, and $rep_i$ the number of Republican hashtags.
We measure the \emph{political polarization} as the normalized difference between the number of Republican and Democrat hashtags used:
\begin{equation}
	\mathcal{P}(u_i) = \frac{rep_i - dem_i}{rep_i + dem_i}.
\end{equation}
$\mathcal{P}(u_i)$ takes values between $-1$ (if $u_i$ emitted only Democrat partisan hashtags) and $1$ ($u_i$ emitted only Republican hashtags).
We threshold the political polarization to construct a population of Democrat users with $\mathcal{P}(u) \leq -0.4$ and Republican users with $\mathcal{P}(u) \geq 0.4$.
In the set of politically engaged users, there are 21,711 Democrat users, 22,644 Republican users and 2,551 users with no polarization ($\mathcal{P}(u) \in (-0.4, 0.4)$).
We measure the \emph{political engagement} of users using the total volume of partisan hashtags included in their tweets $\mathcal{E}(u_i) = rep_i + dem_i$.

\subsection{Botness score $\zeta$ and bot detection}
\label{subsec:bot-detection}


\textbf{Detecting automated bots.}
We use the BotOrNot~\cite{davisetal.16} API to measure the likelihood of a user being a bot for each of the 1,451,388 users in the \debate dataset.
Given a user $u$, the API returns the botness score $\zeta(u) \in [0, 1]$ (with 0 being likely human, and 1 likely non-human).
Previous work~\cite{varol.17,FM7090,Woolley.2017} use a botness threshold of $0.5$ to detect socialbots.
However, we manually checked a random sample of 100 users with $\zeta(u) > 0.5$ and we found several human accounts being classified as bots.
A threshold of 0.6 decreases mis-classification by $3\%$.
It has been previously reported by \citet{varol.17} that organizational accounts have high botness scores.
This however is not a concern in this work, as we aim to detect 
`highly automated' accounts that behave in a non-human way. 
We chose to use a threshold of $0.6$ to construct the \Bot population in light of the more encompassing notion of account \emph{automation}. 
\begin{table}[tb]
	\renewcommand{\cellalign}{cr}
	\small
	\setlength{\tabcolsep}{4pt}
	\centering
	\caption{
		Tabulating population volumes and percentages of politically polarized users over four populations: \Protected, \Human, \Suspended and \Bot.
	}
	\begin{tabular}{rrrrrr}
		\toprule
			& All & \texttt{Prot.} & \Human & \texttt{Susp.} & \Bot \\
  		\midrule		
			All & 1,451,388 & 45,316 & 499,822 & 10,162 & 17,561 \\
			Polarized & 44,299 & 1,245 & 11,972 & 265 & 435 \\
			Democrat & 21,676 & 585 & 5,376 & 111 & 185 \\
			Republican & 22,623 & 660 & 6,596 & 154 & 250 \\
		\midrule
			Dem. \% & 48.93\% & 46.99\% & 44.90\% & 41.89\% & 42.53\%\\
			Rep. \% & 51.07\% & 53.01\% & 55.10\% & 58.11\% & 57.47\%\\ 
   		\bottomrule
	\end{tabular}
	\label{tab:populations-effectives}
\end{table}

\textbf{Four reference populations.}
In addition to the \Bot population, we construct three additional reference populations:
\Human $\zeta(u) \leq 0.2$ contains users with a high likelihood of being regular Twitter users.
\Protected are the users whose profile has the access restricted to their followers and friends (the BotOrNot system cannot return the botness score); we consider these users to be regular Twitter users, since we assume that no organization or broadcasting bot would restrict access to their profile.
\emph{Suspended} are those users which have been suspended by Twitter between the date of the tweet collection (26 September 2016) and the date of retrieving the botness score (July 2017);
this population has a high likelihood of containing bots.
Table~\ref{tab:populations-effectives} tabulates the size of each population, split over  political polarization.


\section{Evaluation of user influence estimation}
\label{sec:evaluation-influence}


\begin{figure}[tb]
	\centering
	\newcommand\myheight{0.171}
	\subfloat[] {
		\includegraphics[height=\myheight\textheight]{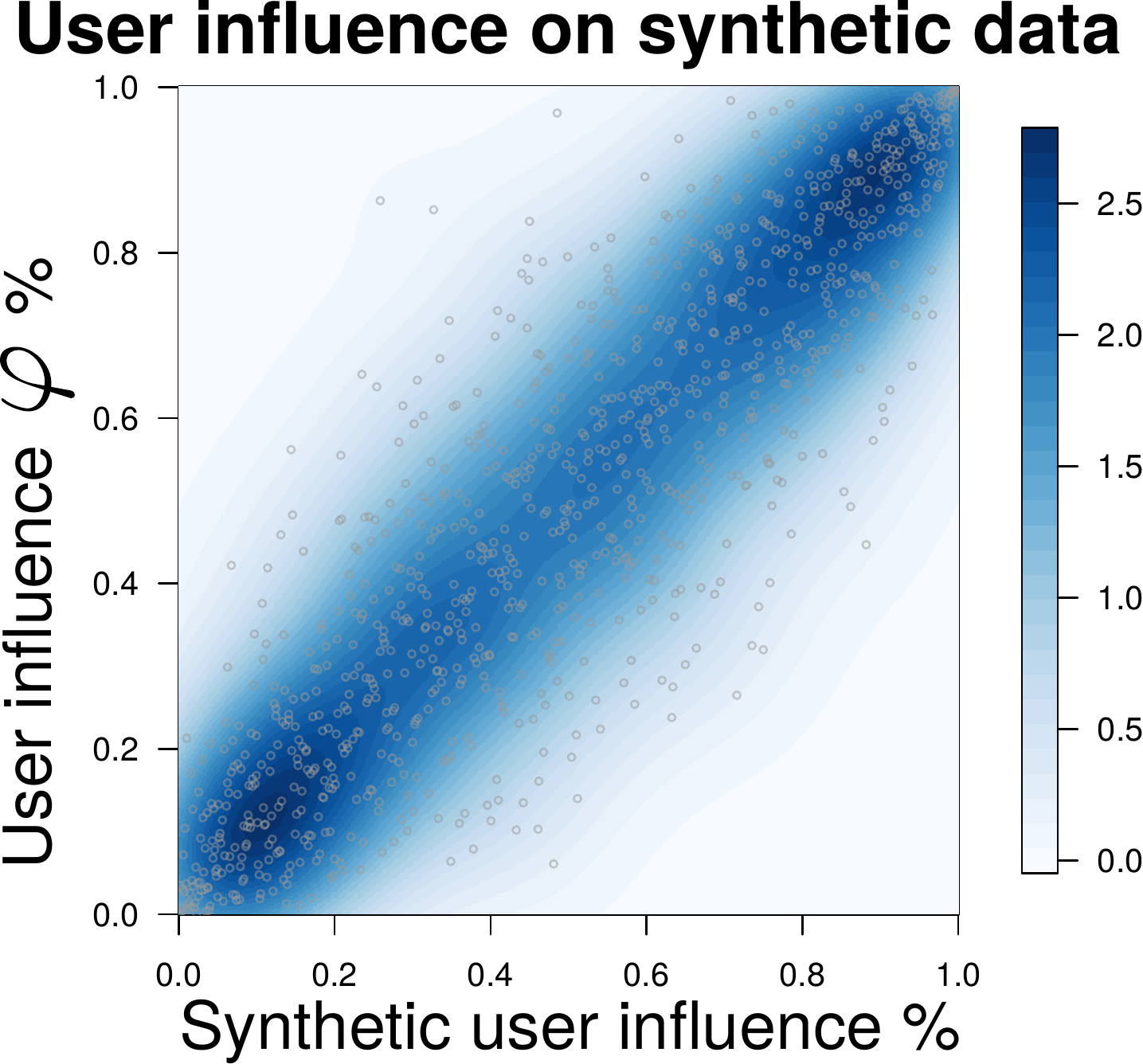}
		\label{subfig:artificial-compare}
	}
	\subfloat[] {
		\includegraphics[height=\myheight\textheight]{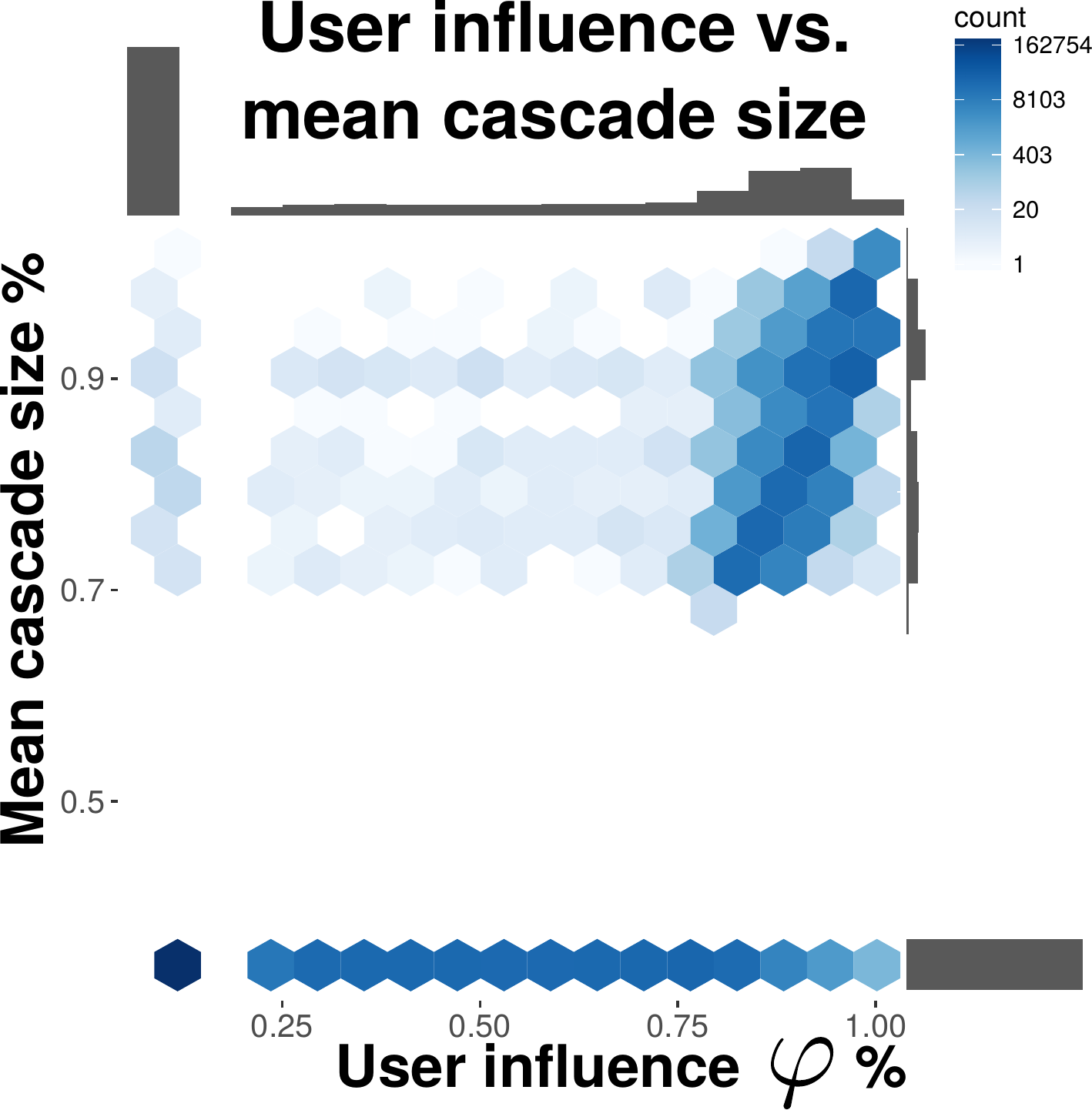}
		\label{subfig:mean-cascade-size}
	}\\
	\subfloat[] {
		\includegraphics[height=0.19\textheight]{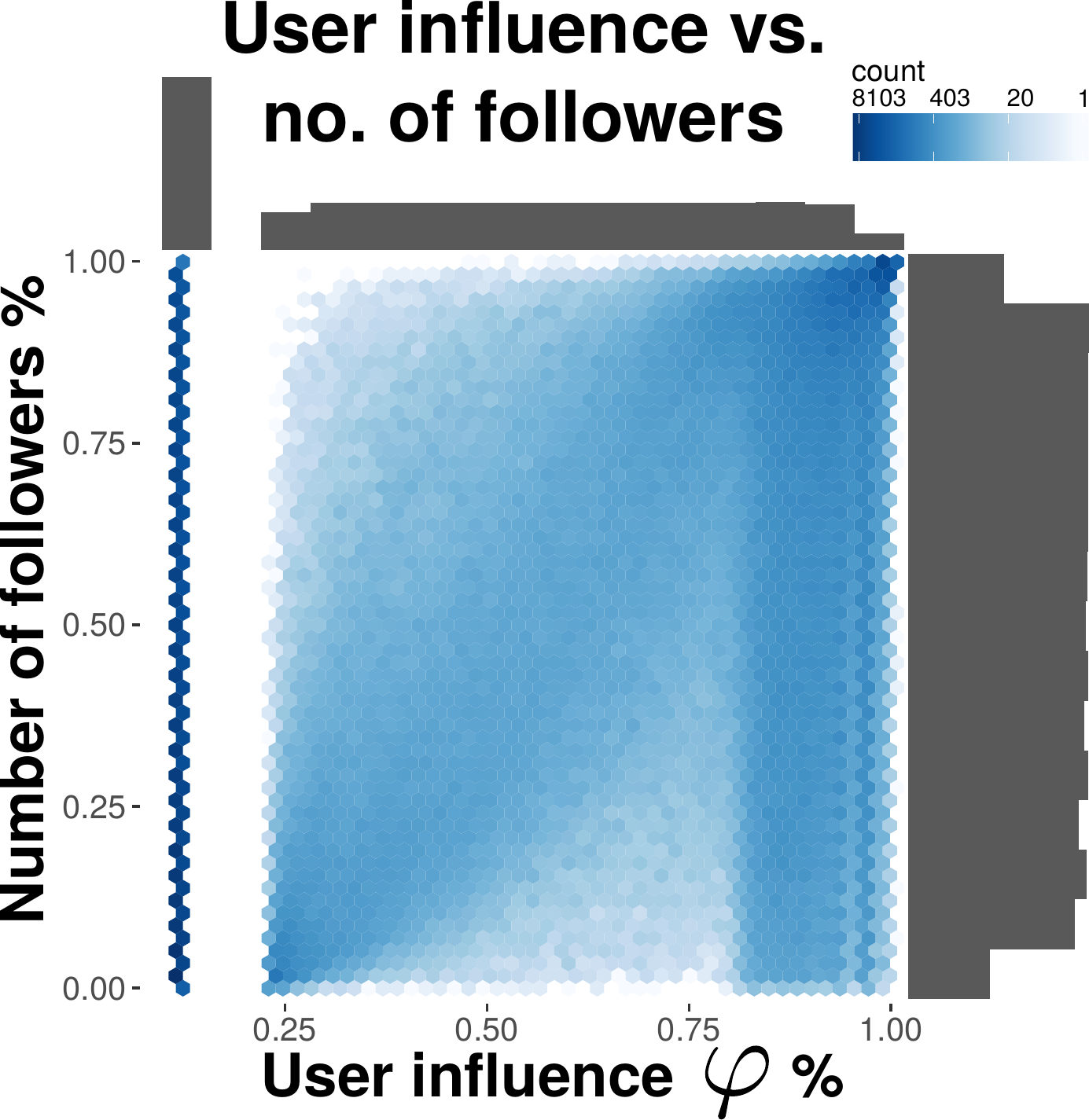}
		\label{subfig:no-followers}
	}
	\caption{ 
		Evaluation of the user influence measure.
		\textbf{(a)} 2D density plot (shades of blue) and scatter-plot (gray circles) of user influence against the ground truth on a synthetic dataset.
		\textbf{(b)(c)} Hexbin plot of user influence percentile (x-axis) against mean cascade size percentile (b) and the number of followers (c) (y-axis) on \debate.
		The color intensity indicates the number of users in each hex bin.
		1D histograms of each axis are shown using gray bars.
		Note $72.3\%$ of all users that initiate cascades are never retweeted.
	}
\end{figure}


\begin{figure*}[htbp]
	\centering
	\newcommand\myheight{0.131}
	\subfloat[] {
		\includegraphics[height=\myheight\textheight]{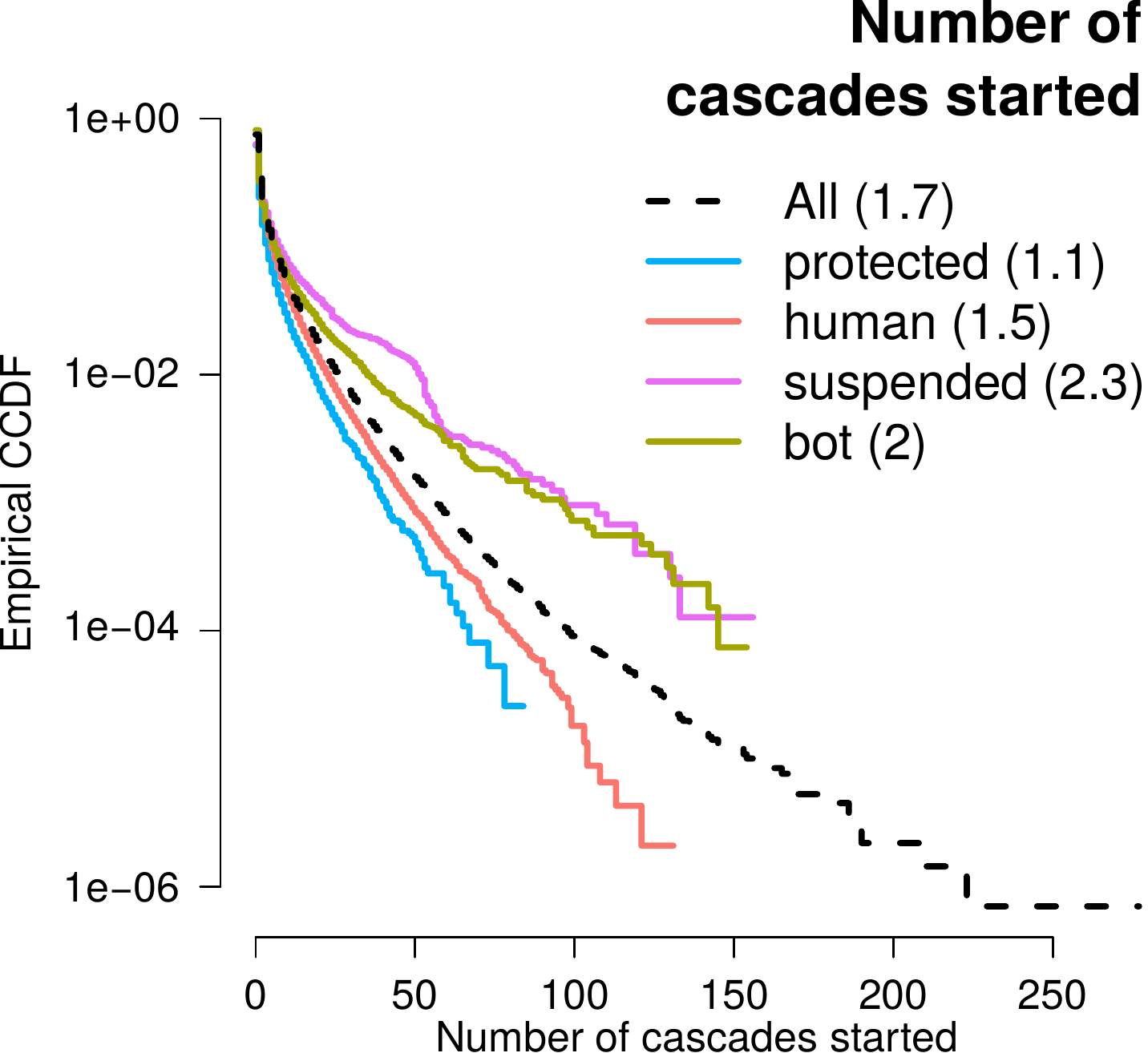}
		\label{subfig:no-cascades}
	}
	\subfloat[] {
		\includegraphics[height=\myheight\textheight]{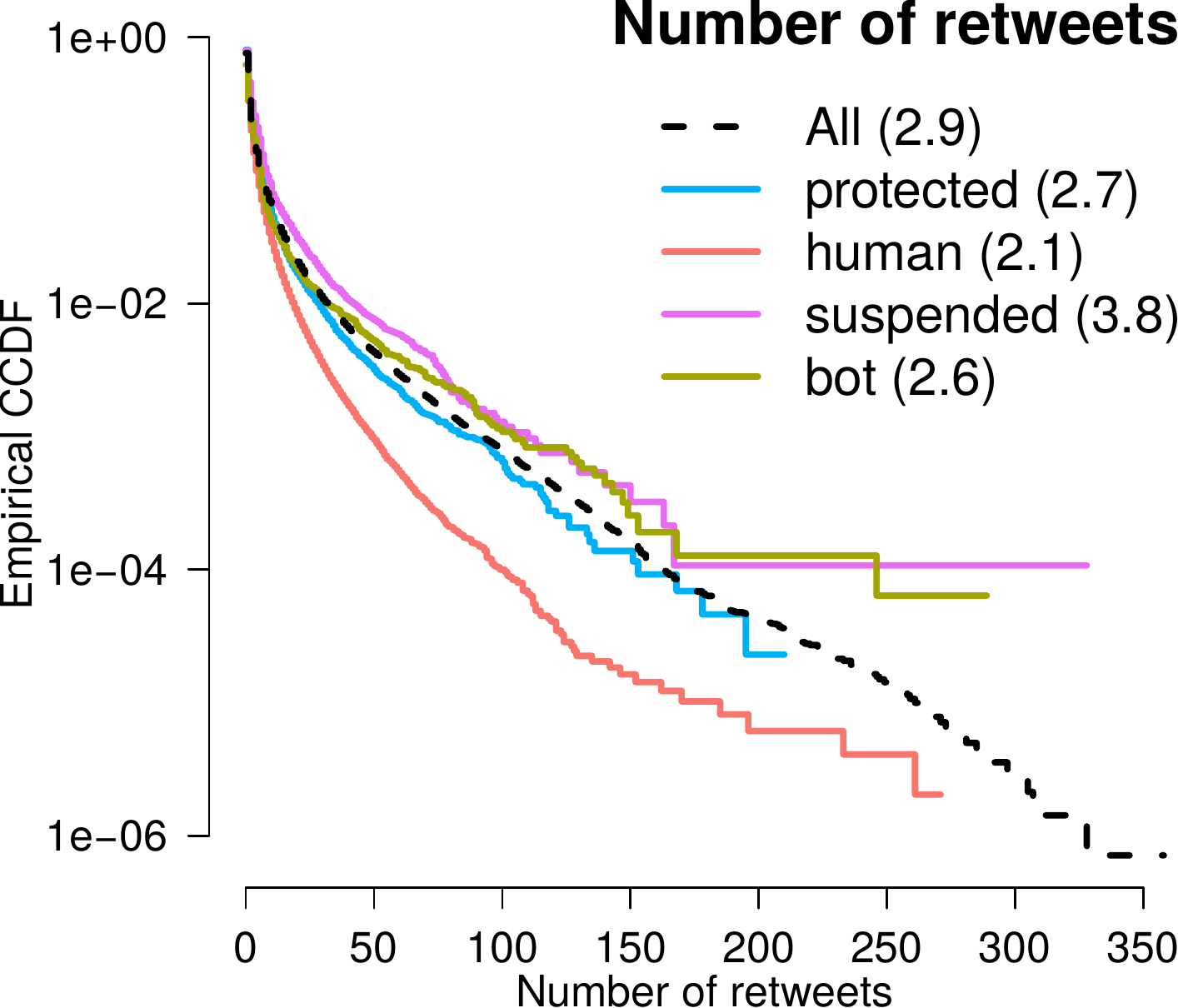}
		\label{subfig:no-retweets}
	}
	\subfloat[] {
		\includegraphics[height=\myheight\textheight]{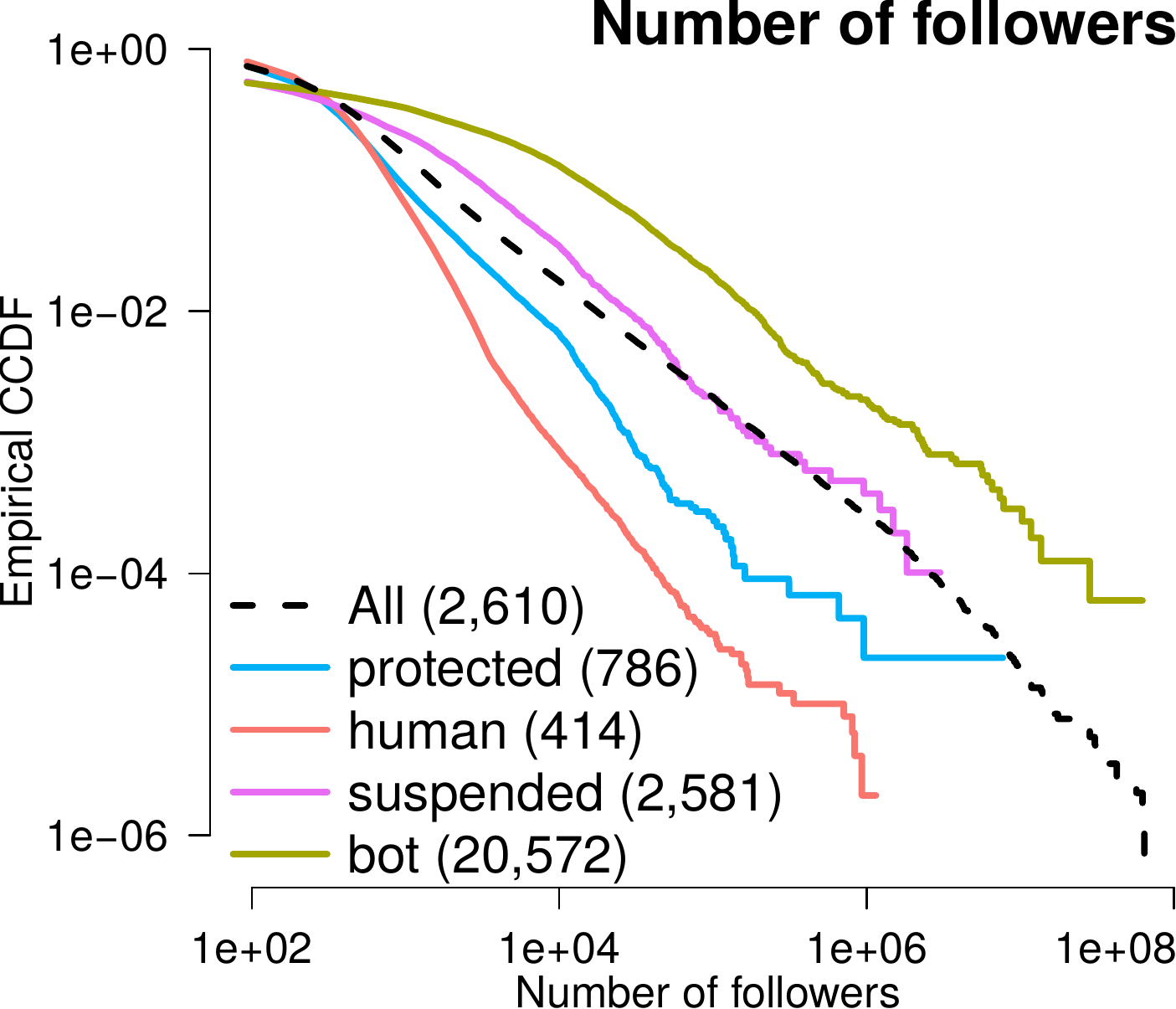}
		\label{subfig:numfolowers-CCDF}
	}
	\subfloat[] {
		\includegraphics[height=\myheight\textheight]{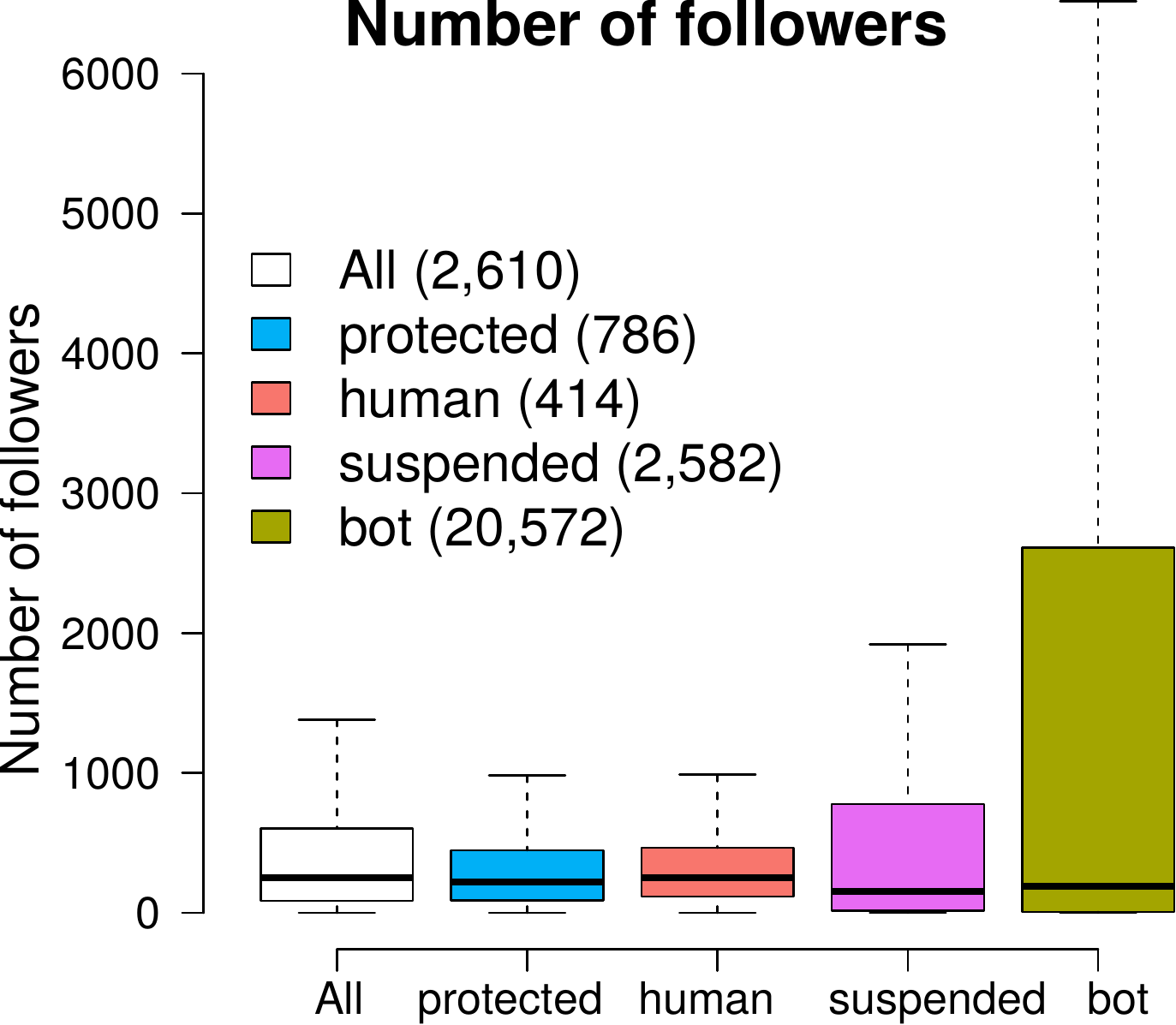}
		\label{subfig:numfolowers-boxplot}
	}
	\subfloat[] {
		\includegraphics[height=\myheight\textheight]{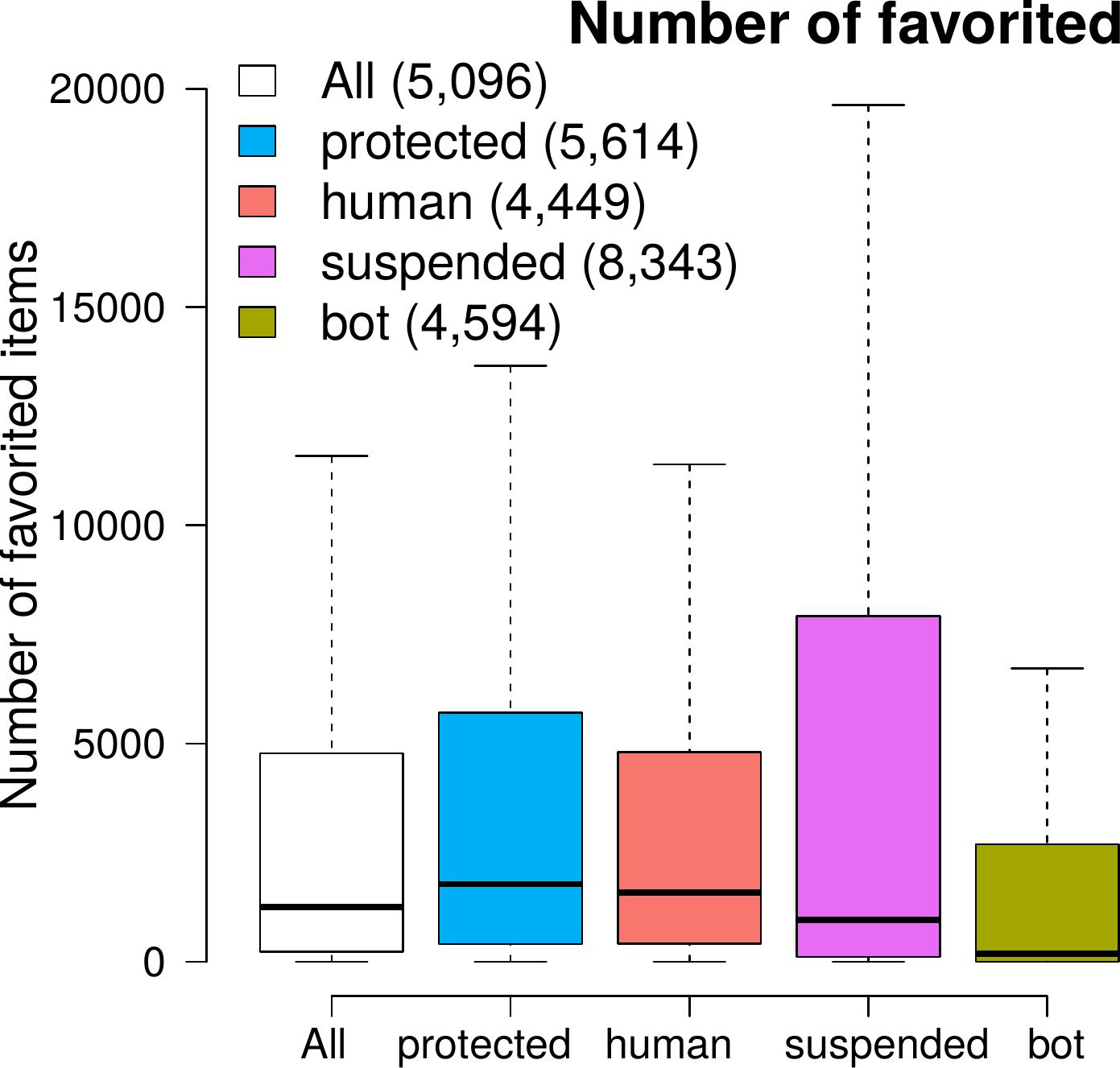}
		\label{subfig:numfavorited}
	}
	\caption{ 
		Profiling behavior of the \Protected, \Human, \Suspended and \Bot populations in the \debate dataset.
		The numbers in parentheses in the legend are mean values.
		\textbf{(a)} CCDF of the number of Twitter diffusion cascades started.
		\textbf{(b)} CCDF of the number of retweets. 
		\textbf{(c)(d)} CCDF (c) and boxplots (d) of the number of followers. 
		\textbf{(e)} Number of items favorited.
	}
	\label{fig:bot-profiling}
\end{figure*}


In this section, we evaluate our proposed algorithm and measure of user influence.
In Sec~\ref{subsec:ground-truth}, we evaluate on synthetic data against a known ground truth.
In Sec.~\ref{subsec:two-alternatives}, we compare the $\varphi(u)$ measure (defined in Sec.~\ref{subsec:user-influence-measure}) against two alternatives: the number of followers and the mean size of initiated cascades.

\subsection{Evaluation of user influence}
\label{subsec:ground-truth}

Evaluating user influence on real data presents two major hurdles.
The first is the lack of ground truth, as user influence is not directly observed.
The second hurdle is that the diffusion graph is unknown, which renders impossible comparing to state-of-the-art methods which require this information (e.g. ConTinEst~\cite{Du2013}).
In this section, evaluate our algorithm against a known ground truth on a synthetic dataset, using the same evaluation approach used for ConTinEst.

\textbf{Evaluation on synthetic data.}
We evaluate on synthetic data using the 
protocol previously employed in~\cite{Du2013}.
We use the simulator in \cite{Du2013} to generate an artificial social network with 1000 users.
We then simulate 1000 cascades through this social network, starting from the same initial user.
The generation of the synthetic social network and of the cascades is detailed in the online supplement~\cite[annex~C]{supplemental}.
Similar to the retweet cascades in \debate, each event in the synthetic cascades has a timestamp and an associated user.
Unlike the real retweet cascades, we know the real diffusion structure behind each synthetic cascade.
For each user $u$, we count the number of nodes reachable from $u$ in the diffusion tree of each cascade.
We compute the influence of $u$ as the mean influence over all cascades.
ConTinEst~\cite{Du2013} has been shown to asymptotically approximate this synthetic user influence.

We use our algorithm introduced in Sec.~\ref{subsec:user-influence-mt} on the synthetic data, to compute the measure $\varphi(u)$ defined in Eq.~\ref{eq:user-infl-casin}.
We plot in Fig.~\ref{subfig:artificial-compare} the 2D scatter-plot and the density plot of the synthetic users, with our influence measure $\varphi$ on the y-axis and the ground truth on the x-axis (both in percentiles).
Visibly, there is a high agreement between the two measures, particularly for the most influential and the least influential users.
The Spearman correlation coefficient of the raw values is $0.88$.
This shows that our method can output reliable user influence estimates in the absence of any information about the structure of the diffusions.

\subsection{Comparison with other influence metrics}
\label{subsec:two-alternatives}

We compare the influence measure $\varphi(u)$ against two alternatives that can be computed on \debate.


\textbf{Mean size of initiated cascades} (of a user $u$) is the average number of users reached by original content authored by $u$.
It should be noted that this measure does not capture $u$'s role in diffusing content authored by someone else.
In the context of Twitter, mean size of initiated cascades is the average number of users who retweeted an original tweet authored by $u$: we compute this for every user in the \debate dataset, and we plot it against $\varphi(u)$ in Fig.~\ref{subfig:mean-cascade-size}.
Few users have a meaningful value for mean cascade size: 
$55\%$ of users never start a cascades (and they are not accounted for in Fig.~\ref{subfig:mean-cascade-size}); 
out of the ones that start cascades $72.3\%$ are never retweeted and they are all positioned at the lowest percentile (shown by the 1D histograms in the plot).
%
It is apparent that the mean cascade size metric detects the influential users that start cascades, and it correlates with $\varphi(u)$.
However, it misses highly influential users who never initiate cascades, but who participate by retweeting. Examples are user \textit{@SethMacFarlane} (the actor and filmmaker Seth MacFarlane, 10.8 million followers) or user \textit{@michaelianblack} (comedian Michael Ian Black, 2.1 million followers), both with $\varphi$ in the top $0.01\%$ most influential users.

\textbf{Number of followers} is one of the simplest measures of direct influence used in literature~\cite{Mishra2016,Zhao2015}.
While being loosely correlated with $\varphi(u)$ (visible in Fig.~\ref{subfig:no-followers}, Pearson $r = 0.42$), it has the drawback of not accounting for any of the user actions, such as an active participation in discussions or generating large retweet cascades.
For example, user \textit{@PoliticJames} (alt-right and pro-Trump, 2 followers) emitted one tweet in \debate, which was retweeted 18 times and placing him in the top $1\%$ most influential users.
Similarly, user \textit{@tiwtter1tr4\_tv} (now suspended, 0 followers) initiated a cascade of size 58 (top $1\%$ most influential).
Interestingly, half of the accounts scoring on the bottom $1\%$ by number of followers and top $1\%$ by influence are now suspended or have very high botness scores.



\section{Results and findings}
\label{sec:results-findings}

In this section, we present an analysis of the interplay between botness, political behavior (polarization and engagement) and influence.
In Sec.~\ref{subsec:polarization-botness}, we first profile the activity of users in the four reference populations; next, we analyze the political polarization and engagement, and their relation with the botness measure.
Finally, in Sec.~\ref{subsec:user-influence-results} we tabulate user influence against polarization and botness, and we construct \emph{the polarization map}.


\begin{figure}[!tb]
	\centering
	{
	\newcommand\myheight{0.145}
	\subfloat[] {
		\includegraphics[height=\myheight\textheight]{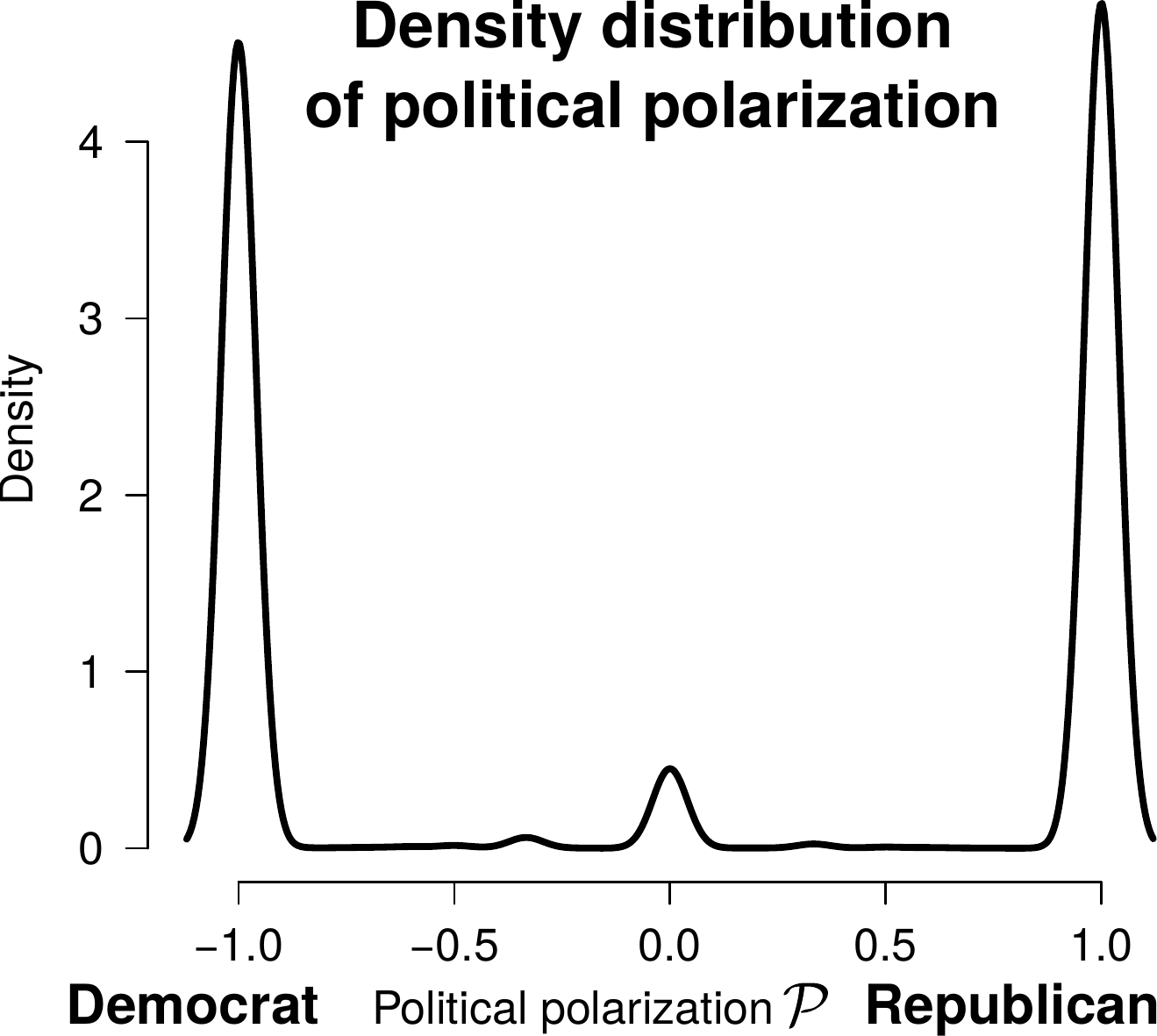}
		\label{subfig:distribution-political-bias}
	}
	\subfloat[] {
		\includegraphics[height=\myheight\textheight]{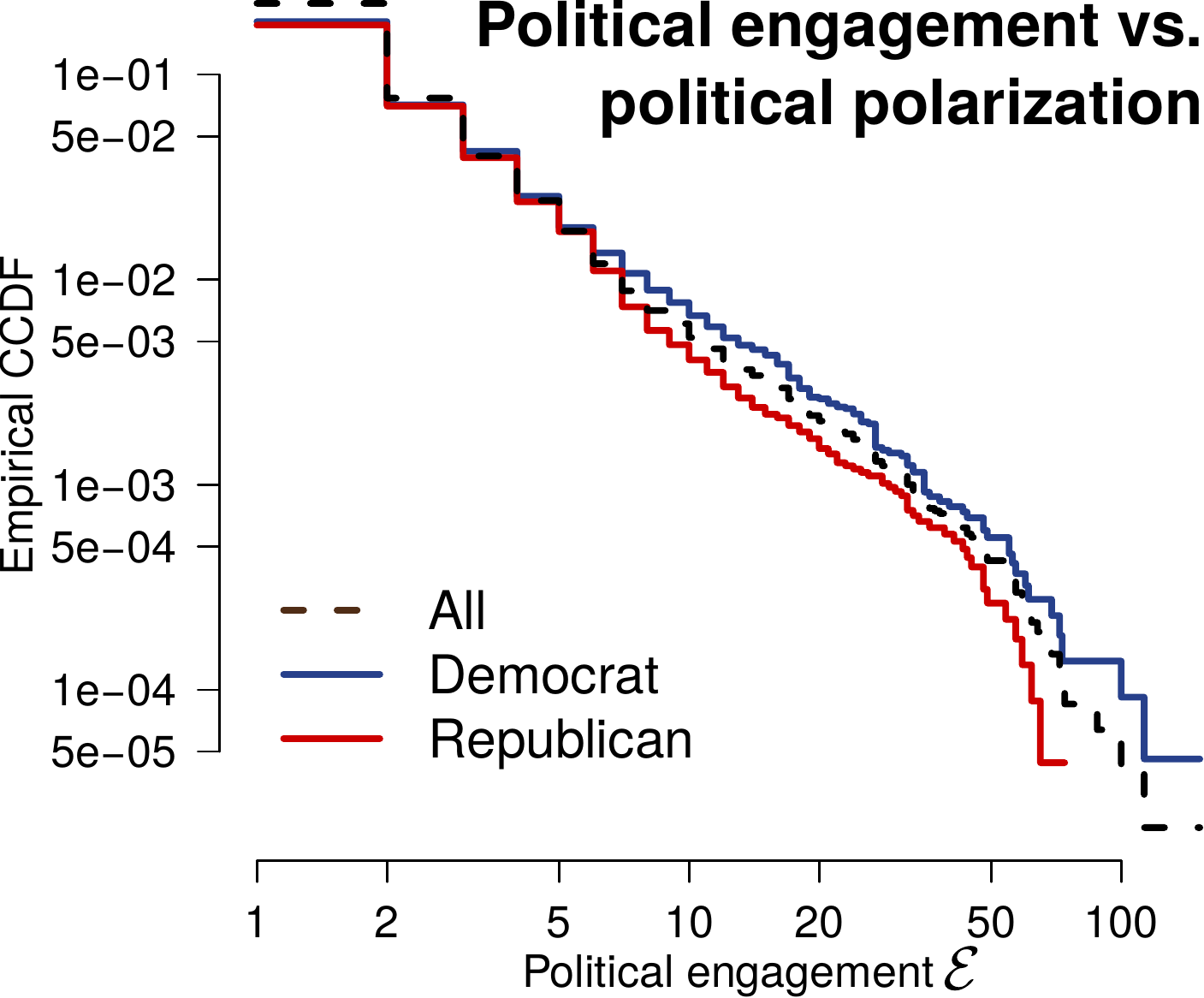}
		\label{subfig:engagement-vs-political}
	}}
	\\
	{
	\newcommand\myheight{0.16}
	\subfloat[] {
		\includegraphics[height=\myheight\textheight]{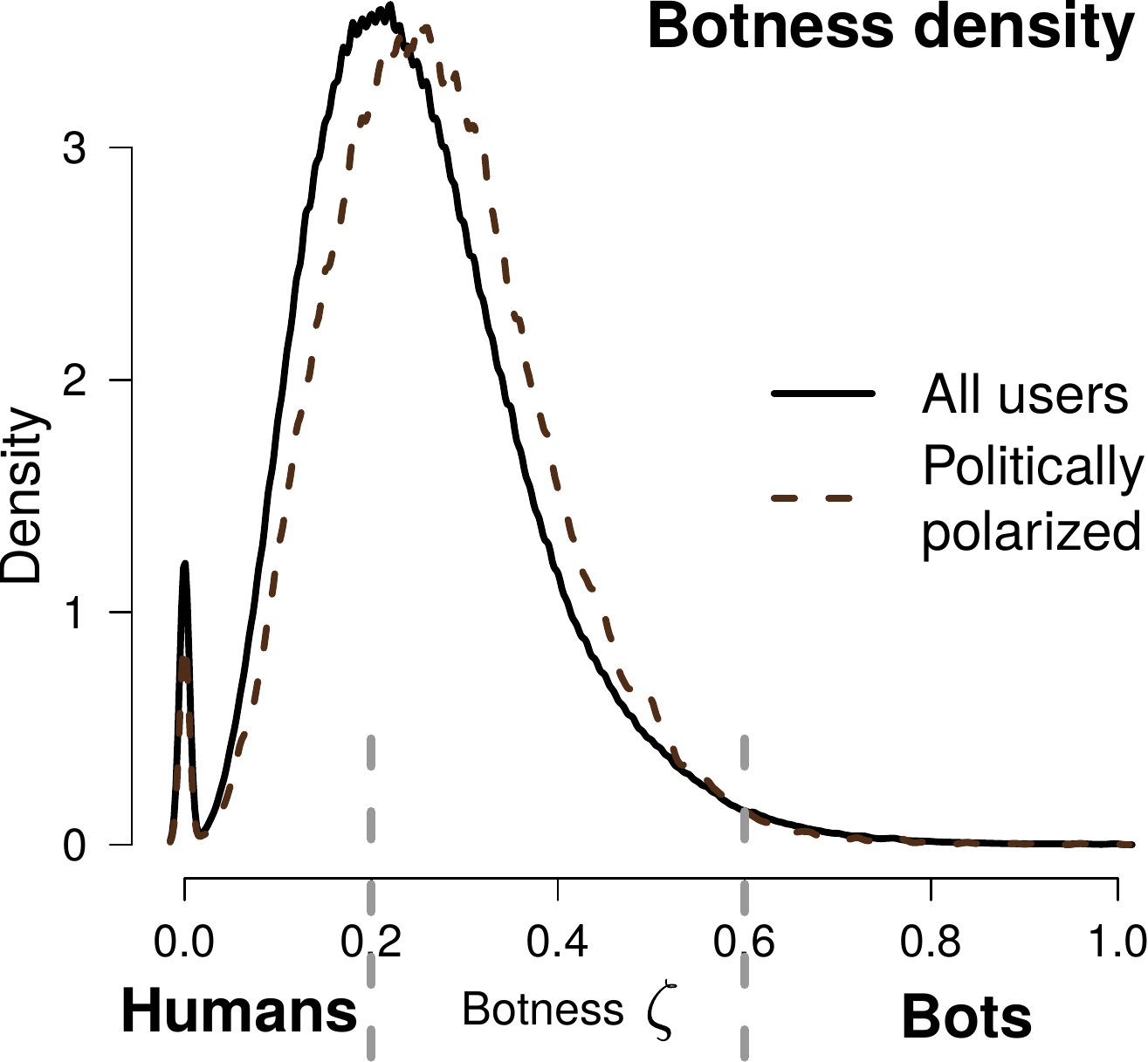}
		\label{subfig:botscore-density}
	}
	\subfloat[] {
		\includegraphics[height=\myheight\textheight]{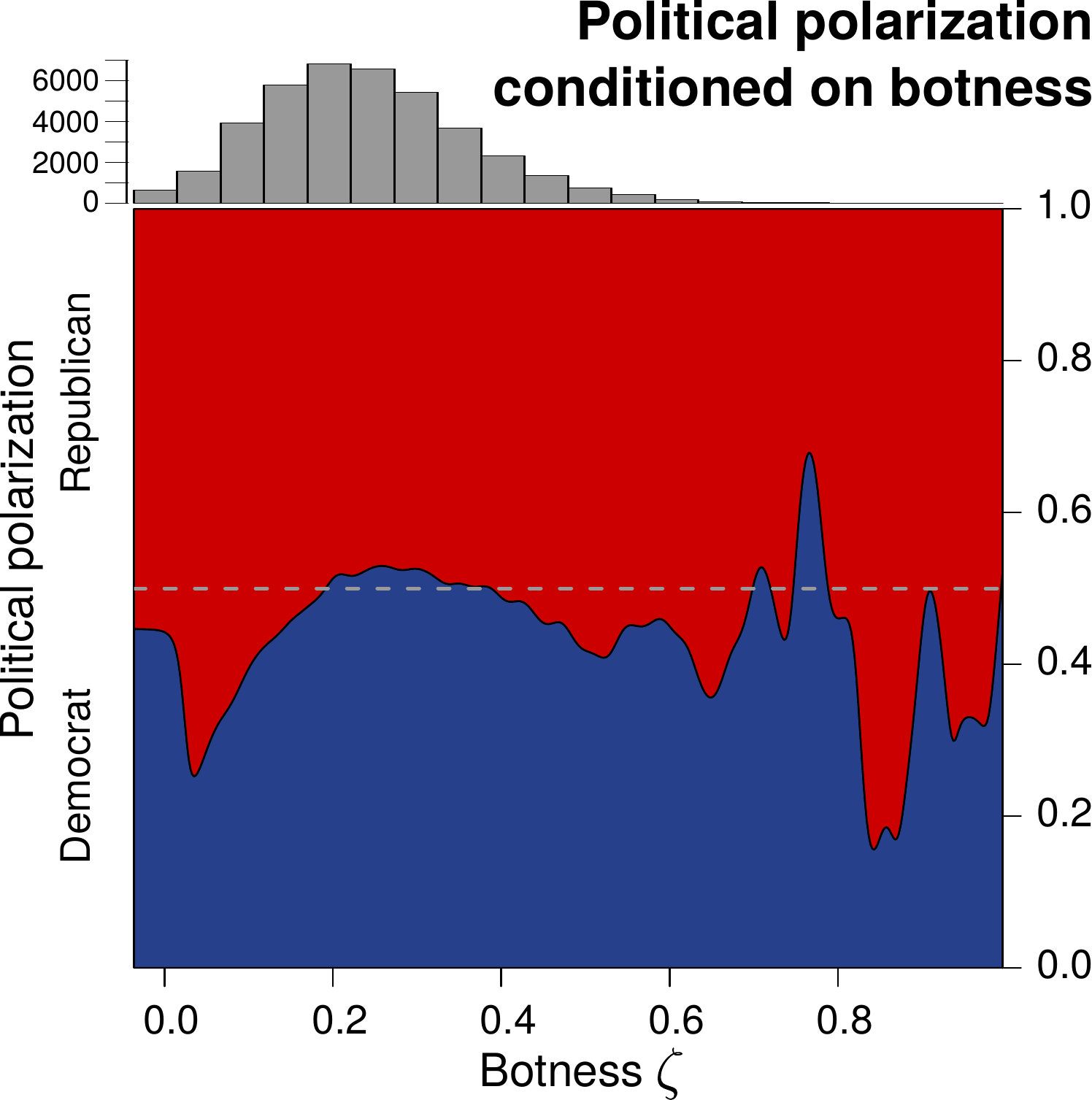}
		\label{subfig:botscore-conditional-density}
	}}
	\\
	{ 
	\newcommand\myheight{0.145}
	\subfloat[] {
		\includegraphics[height=\myheight\textheight]{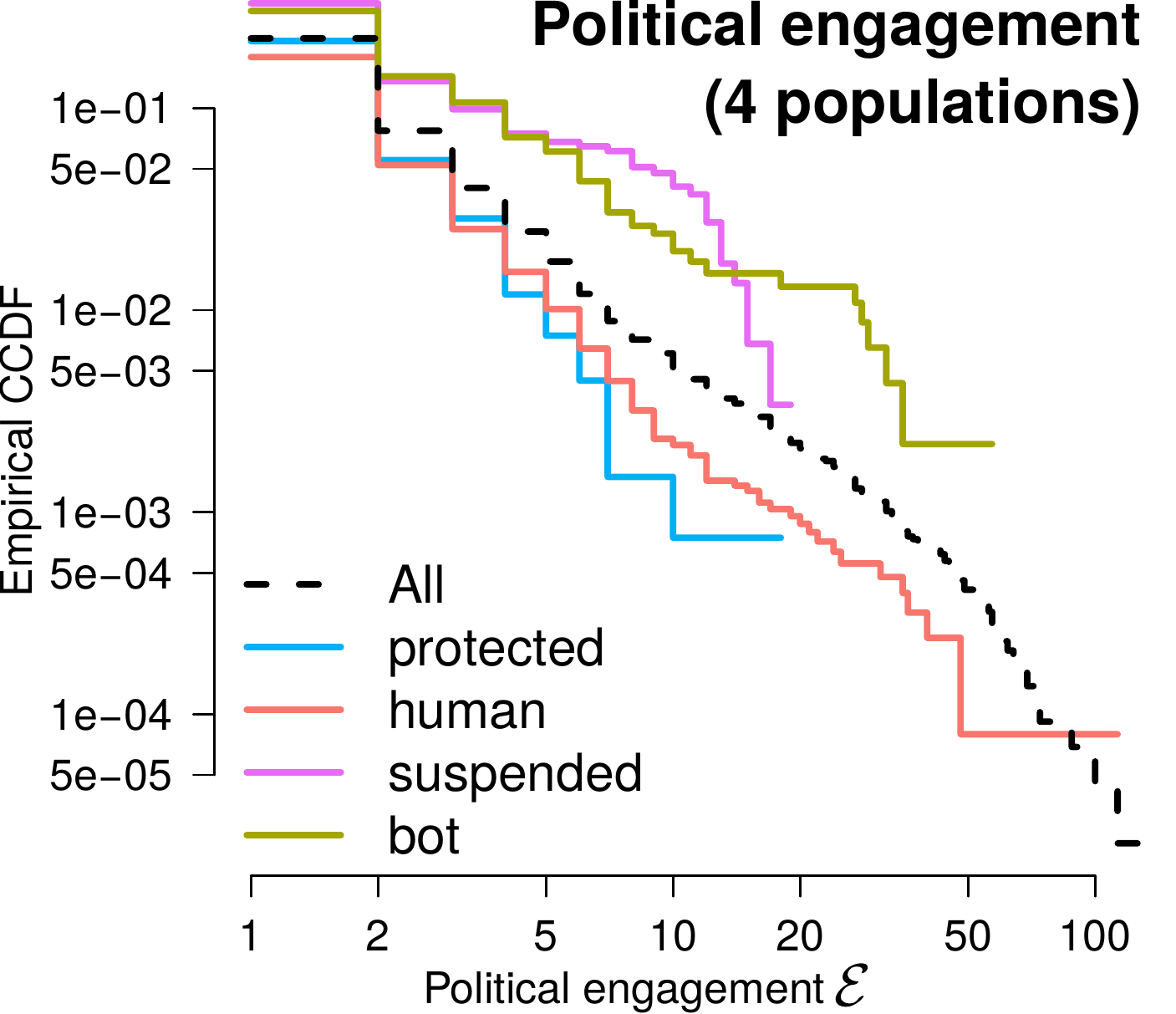}
		\label{subfig:engagement-vs-botscore}
	}
	\subfloat[] {
		\includegraphics[height=\myheight\textheight]{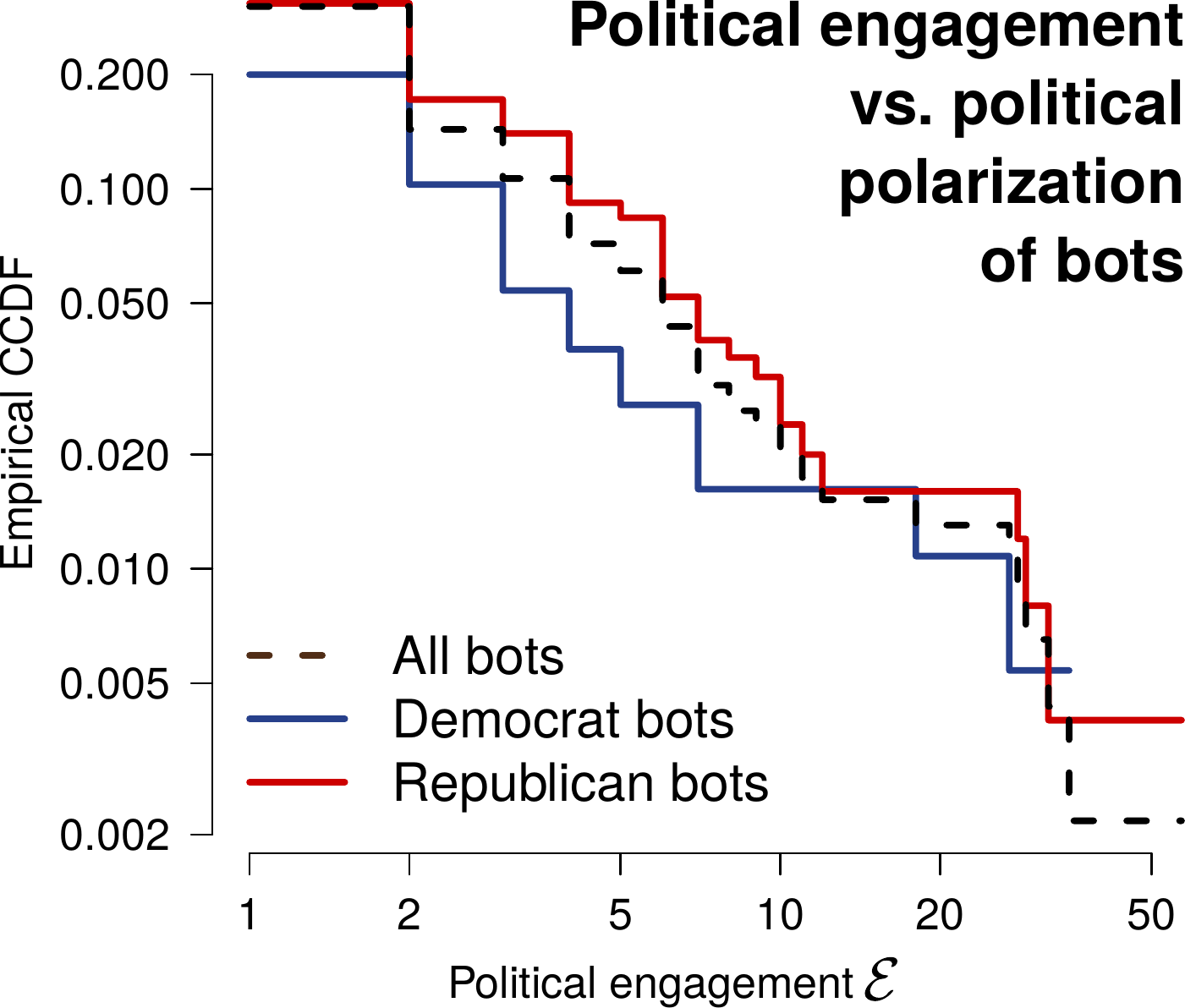}
		\label{subfig:BOTS_pol-engagement-vs-botscore}
	}
	}
	\caption{ 
		Political polarization, engagement and botness.
		\textbf{(a)} The density distribution of political polarization $\mathcal{P}$. 
		\textbf{(b)} Log-log plot of the CCDF of political engagement $\mathcal{E}$ 
		for the Democrat and Republican populations.
		\textbf{(c)} The density distribution of botness $\zeta$ for the entire population (solid line) and the politically polarized population (dashed line). 
		\textbf{(d)} The conditional density of polarization conditioned on botness.
		The top panel shows the volumes of politically polarized users in 30 bins.
		\textbf{(e)(f)} CCDF of political engagement for the reference populations (e) and for the polarized \Bot populations (f).
	}
	\label{fig:bot-polarization}
\end{figure}

\subsection{Political behavior of humans and bots}
\label{subsec:polarization-botness}

\textbf{Twitter activity across four populations.}
We measure the behavior of users in the four reference populations defined in Sec.~\ref{subsec:political-polarization-measures} using several measures computed from the Twiter API.
The number of cascades started (i.e., number of original tweets) and the number of posted retweets are simple measures of activity on Twitter, and they are known to be long-tail distributed~\cite{Cha2010}.
Fig.~\ref{subfig:no-cascades} and~\ref{subfig:no-retweets} respectively plot the log-log plot of the empirical Complementary Cumulative Distribution Function (CCDF) for each of the two measures.
It is apparent that users in the \Bot and \Suspended populations exhibit higher levels of activity than the general population, whereas the \Human and \Protected populations exhibit lower level.
Fig.~\ref{subfig:numfolowers-CCDF} and~\ref{subfig:numfolowers-boxplot} plot the number of followers and present a more nuanced story:
the average bot user has 10 times more followers than the average human user;
however, bots have a median of $190$ followers, less than the median $253$ followers of human users.
In other words, some bots are very highly followed, but most are simply ignored.
Finally, Fig.~\ref{subfig:numfavorited} shows that bots favorite less than humans, indicating that their activity patterns differ from those of humans.

\textbf{Political polarization and engagement.}
The density distribution of political polarization (Fig.~\ref{subfig:distribution-political-bias}) shows two peaks at -1 and 1, corresponding to strongly pro-Democrat and strongly pro-Republican respectively. 
The shape of the density plot is consistent with the sizes of Republican and Democrat populations (Sec.~\ref{subsec:political-polarization-measures}), and the extreme bi-modality can be explained by the clear partisan nature of the chosen hashtags and by the known political polarization of users on Twitter \cite{conover.2011,barbera.2015}, which will be greatly enhanced in the context of a political debate.
Fig.~\ref{subfig:engagement-vs-political} presents the log-log plot of the CCDF of the political engagement, which shows that the political engagement score is long-tail distributed, with \emph{pro-Democrats slightly more engaged than pro-Republicans overall} (t-test significant, p-val $ = 0.0012$).


\textbf{Botness and political polarization.}
The distribution of botness $\zeta$ exhibits a large peak around $[0.1, 0.4]$ and a long tail (Fig.~\ref{subfig:botscore-density}). 
The dashed gray vertical lines show the threshholds used in Sec.~\ref{subsec:bot-detection} for constructing the reference \Human ($\zeta \in [0, 0.2]$) and \Bot ($\zeta \in [0.6, 1]$) populations.
Fig.~\ref{subfig:botscore-conditional-density} shows the conditional density of polarization conditioned on botness.
For both high botness scores (i.e., bots) and low botness scores (humans) the likelihood of being pro-Republican is consistently higher than that of being pro-Democrat, while users with mid-range botness are more likely to be pro-Democrat.
In other words, \emph{socialbots accounts are more likely to be pro-Republican than to be pro-Democrat}.

\textbf{Political engagement of bots.}
Fig.~\ref{subfig:engagement-vs-botscore} shows the CCDF of political engagement of the four reference populations, and it is apparent that the \Bot and \Suspended populations exhibit consistently higher political engagement than the \Human and \Protected populations. 
Fig.~\ref{subfig:BOTS_pol-engagement-vs-botscore} shows the CCDF of political engagement by the political partisanship of bots and we find that pro-Republican \Bot accounts are more politically engaged than their pro-Democrat counterparts.
In summary, \emph{socialbots are more engaged than humans (p-val = $8.55 \times 10^{-5}$), and pro-Republican bots are more engaged than their pro-Democrat counterparts (p-val = 0.1228)}.


\begin{figure*}[htbp]
	\centering
	\newcommand\myheight{0.162}
	\subfloat[] {
		\includegraphics[height=\myheight\textheight]{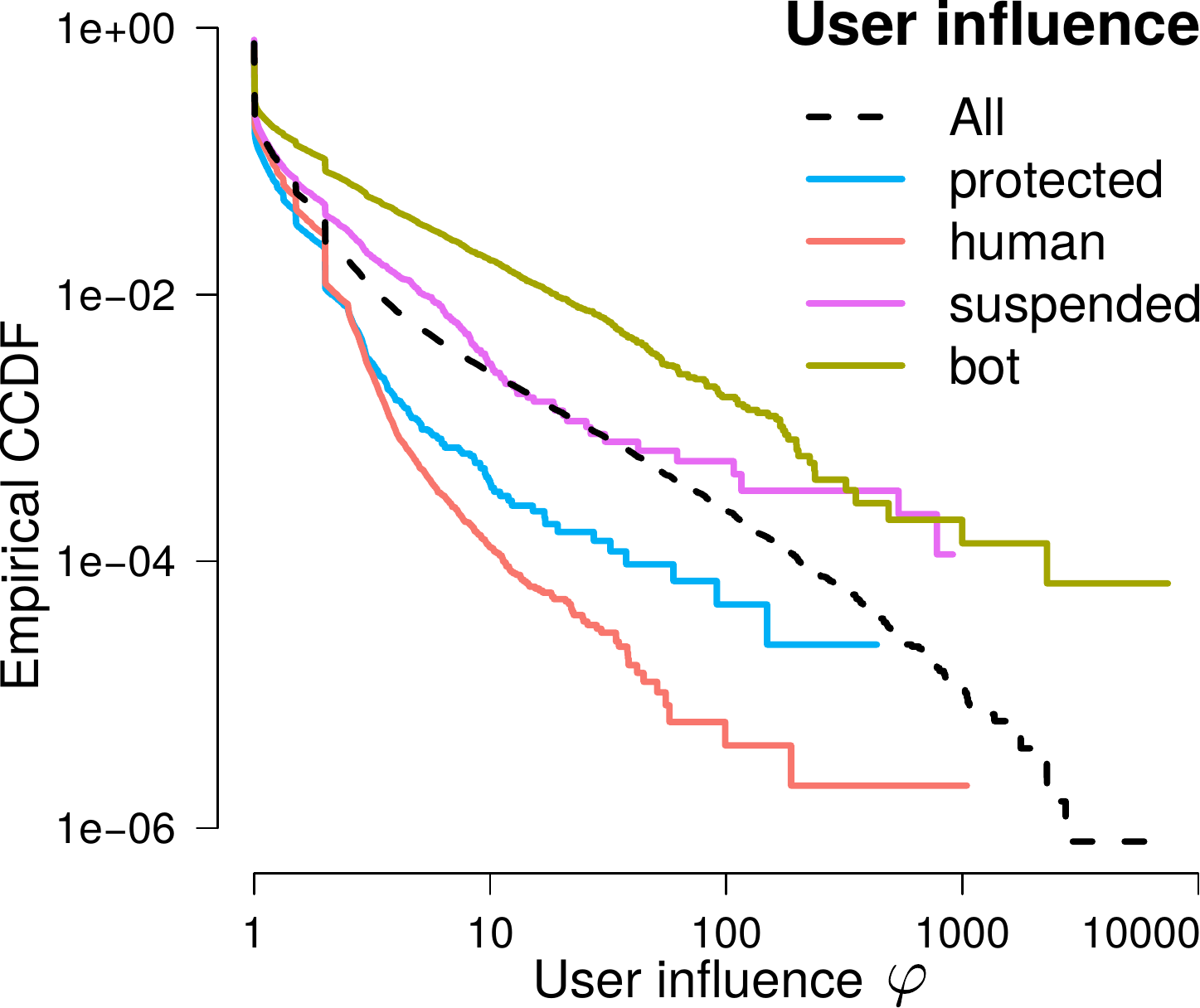}
		\label{subfig:user-infl-CCDF}
	}
	\subfloat[] {
		\includegraphics[height=\myheight\textheight]{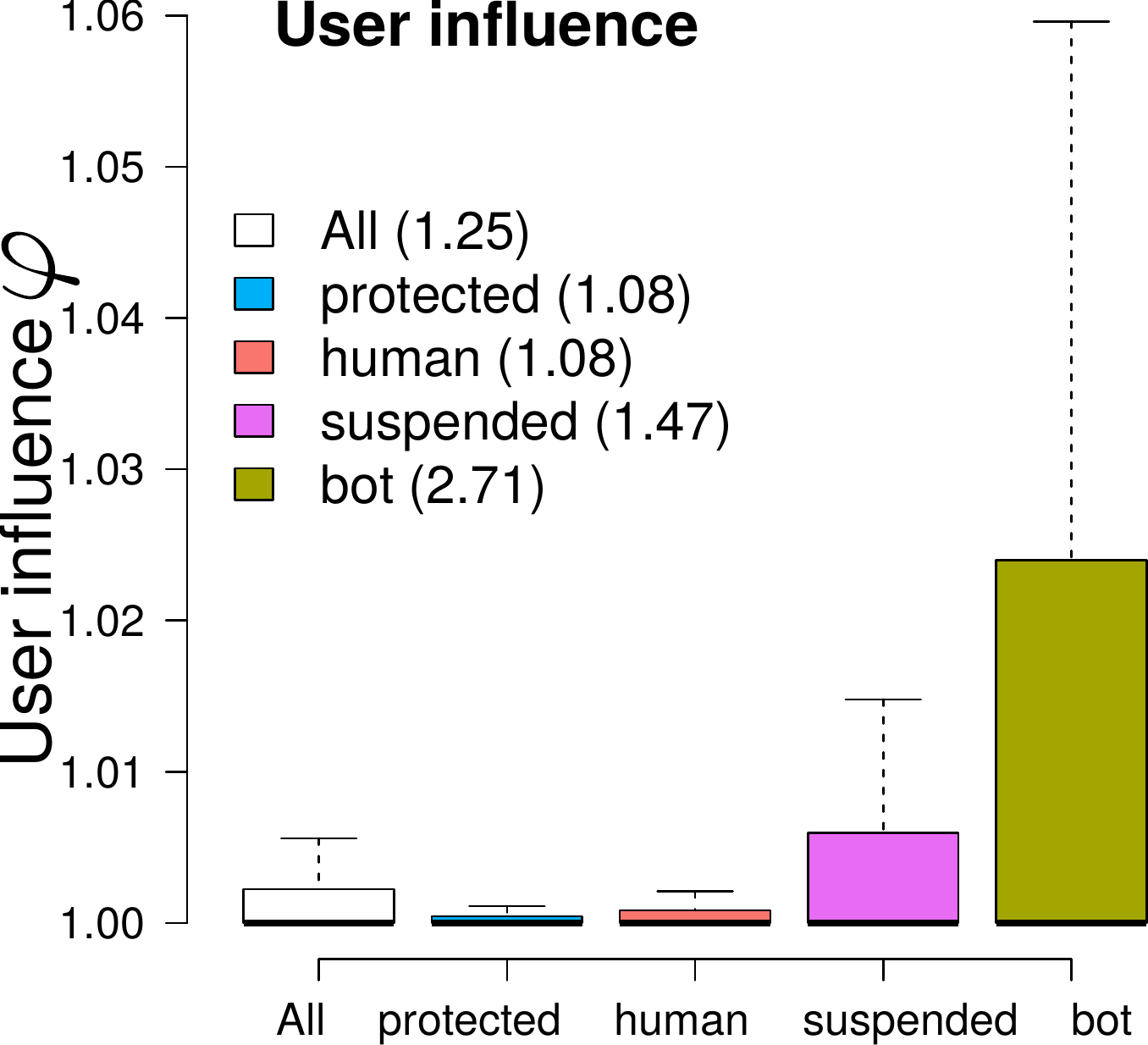}
		\label{subfig:user-infl-boxplots}
	}
	\subfloat[] {
		\includegraphics[height=\myheight\textheight]{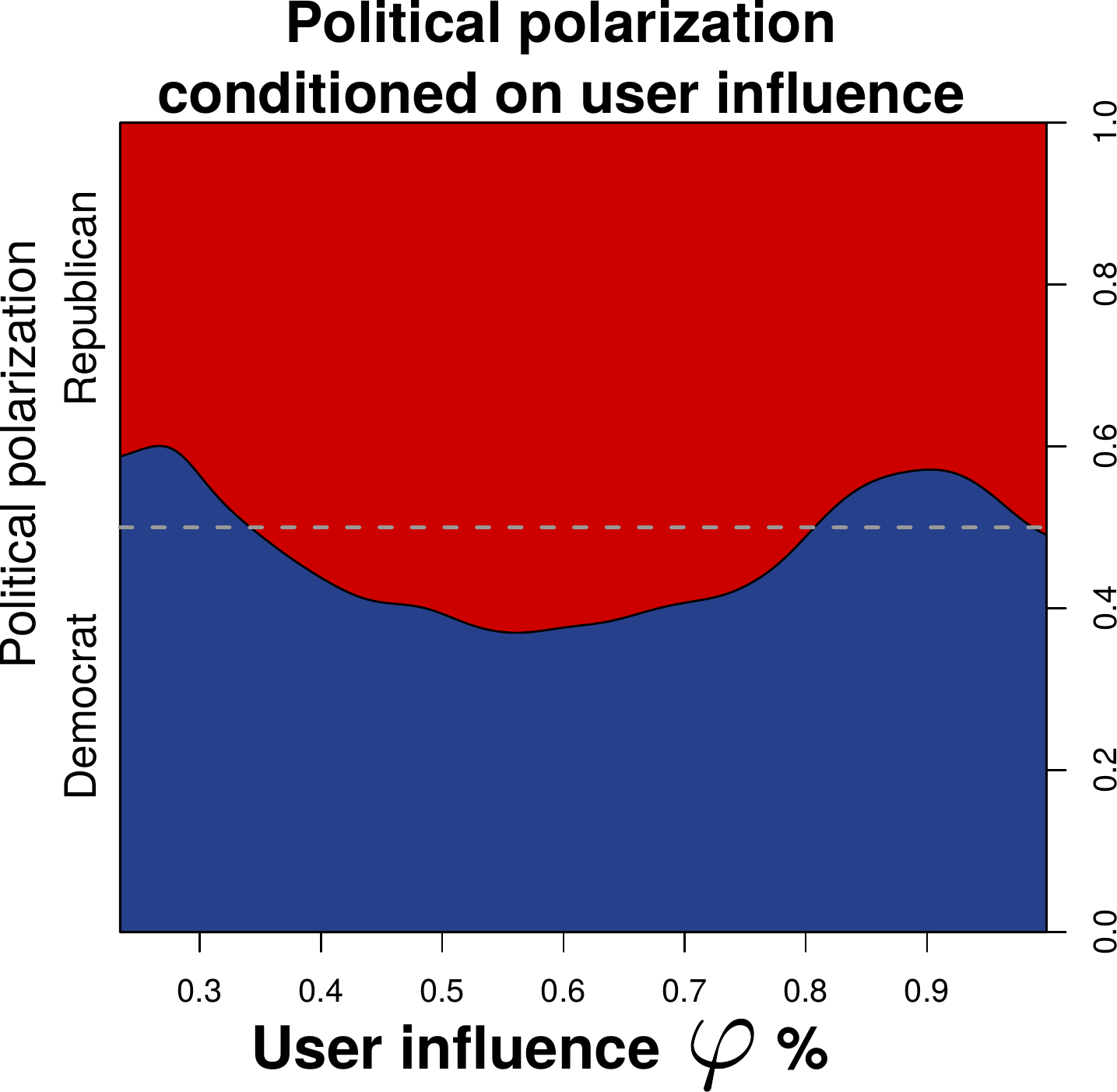}
		\label{subfig:polarization-conditioned-user-infl}
	}
	\subfloat[] {
		\includegraphics[height=\myheight\textheight]{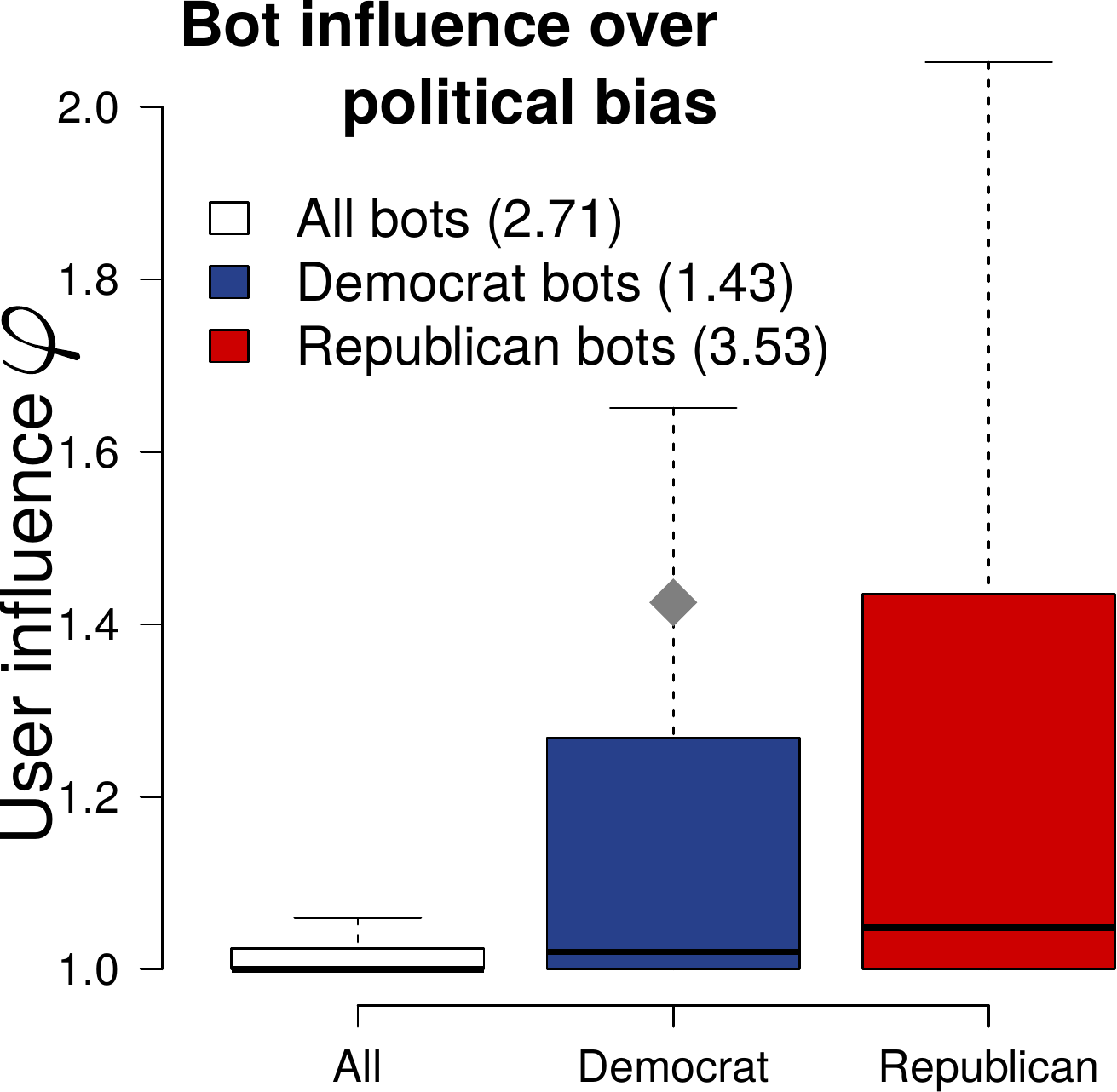}
		\label{subfig:bot-influence}
	}
	\caption{ 
		Profiling influence, and linking to botness and political behavior.
		\textbf{(a)(b)} User influence $\varphi(u)$ for the reference populations, shown as log-log CCDF plot (a) and boxplots (b).
		\textbf{(c)} Probability distribution of polarization, conditional on $\varphi(u) \%$.
		\textbf{(d)} Boxplots of user influence for the pro-Democrat and pro-Republican \Bot users.
		Numbers in parenthesis show mean values.
	}
\end{figure*}


\begin{figure}[tb]
	\centering
	
	\includegraphics[width=0.47\textwidth]{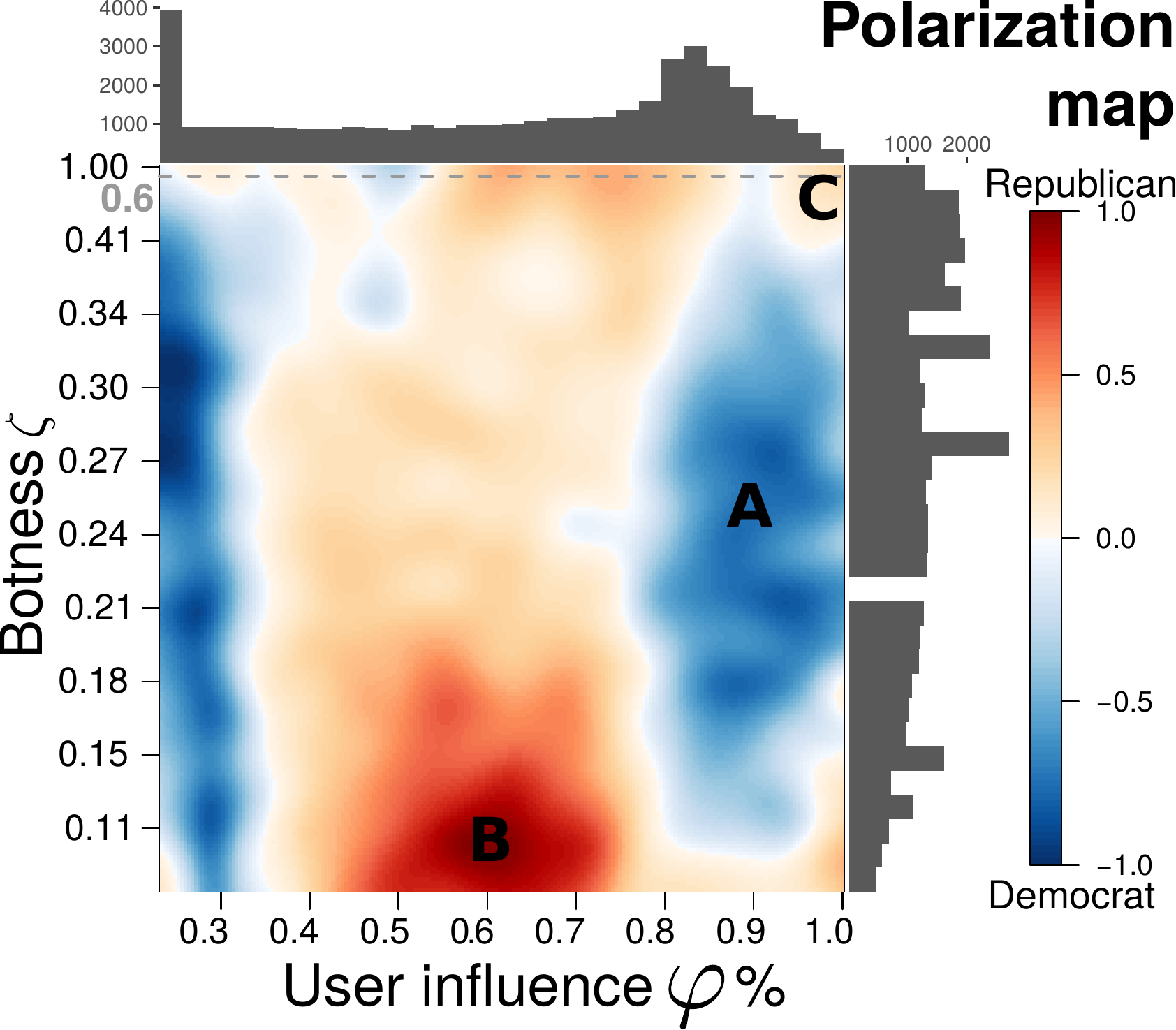}

	\caption{ 
		Political polarization by user influence $\varphi(u) \%$ (x-axis) and bot score $\zeta$ (y-axis).
		The gray dashed horizontal line shows the threshold of 0.6 above which a user is considered a bot.
		The color in the map shows political polarization: areas colored in bright blue (red) are areas where the Democrats (Republicans) have considerably higher density than Republicans (Democrats).
		Areas where the two populations have similar densities are colored white.
		Three areas of interest are shown by the letter \textbf{A}, \textbf{B} and \textbf{C}.
	}
	\label{fig:polarization-map}
\end{figure}

\subsection{User influence and polarization map}
\label{subsec:user-influence-results}

\textbf{User influence across four populations.}
First, we study the distribution of user influence across the four reference populations constructed in Sec.~\ref{subsec:bot-detection}.
We plot the CCDF in Fig.~\ref{subfig:user-infl-CCDF} and we summarize user influence as boxplots in Fig.~\ref{subfig:user-infl-boxplots} for each population.
User influence $\varphi$ is long-tail distributed (shown in Fig.~\ref{subfig:user-infl-CCDF}) and it is higher for \Bot and \Suspended populations, than for \Human and \Protected (shown in Fig~\ref{subfig:user-infl-boxplots}).
There is a large discrepancy between the influence of \Human and \Bot (p-val $= 0.0025$), with \emph{the average bot having 2.5 times more influence than the average human.}
We further break down users in the \Bot population based on their political polarization.
Fig.~\ref{subfig:bot-influence} aggregates as boxplots the influence of pro-Democrat and pro-Republican bots (note: not all bots are politically polarized).
Notably, on a per-bot basis, pro-Republican bots are more influential than their pro-Democrat counterparts (p-val $= 0.0096$) -- \emph{the average pro-Republican bot is twice as influential as the average pro-Democrat bot}.

\textbf{Political polarization and user influence.}
Next, we analyze the relation between influence and polarization.
Fig.~\ref{subfig:polarization-conditioned-user-infl} plots the probability distribution of political polarization, conditioned on user influence $\varphi \%$.
While for mid-range influential users ($\varphi \% \in [0.4, 0.8]$) the likelihood of being Republican is higher than being Democrat, we observe the inverse situation on the higher end of the influence scale.
\emph{Very highly influential users ($\varphi \% > 0.8]$) are more likely to be pro-Democrat}, and this is consistent with the fact that many public figures were supportive of the Democrat candidate during the presidential campaign.

\textbf{The polarization map.}
Finally, we create a visualization that allows us to jointly account for botness and user influence when studying political partisanship.
We project each politically polarized user in \debate onto the two-dimensional space of user influence $\varphi \%$ (x-axis) and botness $\zeta$ (y-axis).
The y-axis is re-scaled so that an equal length interval around any botness value contains the same amount of users,
This allows to zoom in into denser areas like $\zeta \in [0.2, 0.4]$, and to deal with data sparsity around high botness scores.
We compute the 2D density estimates for the pro-Democrat and pro-Republican users (shown in the online supplement~\cite[annex~E]{supplemental}).
For each point in the space $(\varphi \%, \zeta)$ we compute a score as the log of the ratio between the density of the Republican users and that of the pro-Democrats, which is then renormalized so that values range from -1 (mostly Democrat) to +1 (mostly Republican).
The resulting map -- dubbed the \emph{polarization map} -- is shown in Fig.~\ref{fig:polarization-map} and it provides a number of insights.
Three areas of interest (\textbf{A}, \textbf{B} and \textbf{C}) are shown on Fig.~\ref{fig:polarization-map}.
Area \textbf{A} is a pro-Democrat area corresponding to highly influential users (already shown in Fig.~\ref{subfig:polarization-conditioned-user-infl}) that spans across most of the range of botness values.
Area \textbf{B} is the largest predominantly pro-Republican area and it corresponds to mid-range influence (also shown in Fig.~\ref{subfig:polarization-conditioned-user-infl}) and concentrates around small botness values -- this indicates the presence of a large pro-Republican population of mainly human users with regular user influence.
Lastly, we observe that the top-right area \textbf{C} (high botness and high influence) is predominantly red: 
In other words \emph{highly influential bots are mostly pro-Republican.}


\section{Discussion}
 
In this paper, we study the influence and the political behavior of socialbots.
We introduce a novel algorithm for estimating user influence from retweet cascades in which the diffusion structure is not observed.
We propose four measures
to analyze the role and user influence of bots versus humans on Twitter during the 1st U.S. presidential debate of 2016. 
The first is the user influence, computed over all possible unfoldings of each cascade.
Second, we use the BotOrNot API to retrieve the botness score for a large number of Twitter users.
Lastly, by examining the 1000 most frequently-used hashtags we measure political polarization and engagement. 
We analyze the interplay of influence, botness and political polarization using a two-dimensional map -- the polarization map.
We make several novel findings, for example: bots are more likely to be pro-Republican; the average pro-Republican bot is twice as influential as its pro-Democrat counterpart; very highly influential users are more likely to be pro-Democrat; and highly influential bots are mostly pro-Republican.

%
%
%
%
%
\textbf{Validity of analysis with respect to BotOrNot.}
The BotOrNot algorithm uses tweet content and user activity patterns to predict botness.
However, this does not confound the conclusions presented in Sec.~\ref{sec:results-findings}.
First, political behavior (polarization and engagement) is computed from a list of hashtags specific to \debate, while the BotOrNot predictor was trained before the elections took place and it has no knowledge of the hashtags used during the debate.
Second, a loose relation between political engagement and activity patterns could be made, however we argue that engagement is the number of used partisan hashtags, not tweets -- i.e. users can have a high political engagement score after emitting few very polarized tweets.

\textbf{Assumptions, limitations and future work.}
This work makes a number of simplifying assumptions, some of which can be addressed in future work.
First, the delay between the tweet crawling (Sept 2016) and computing botness (July 2017) means that a significant number of users were suspended or deleted.
A future application could see simultaneous tweets and botscore crawling.
Second, our binary hashtag partisanship characterization does not account for independent voters or other spectra of democratic participation, and future work could evaluate our approach against a clustering approach using follower ties to political actors \cite{barbera.2015}.
Last, this work computes the expected influence of users in a particular population, but it does not account for the aggregate influence of the population as a whole.
Future work could generalize our approach to entire populations, which would allow answers to questions like ``Overall, were the Republican bots more influential than the Democrat humans?''.

\vspace{0.2cm}
\noindent{
\textbf{Acknowledgments.}
This research is sponsored in part by the Air Force Research Laboratory, under agreement number FA2386-15-1-4018.
}

{ 
\fontsize{9.0pt}{10.0pt}
\selectfont
\bibliography{paper}
\bibliographystyle{aaai}
}

%
\newpage
\appendix
\etocdepthtag.toc{mtappendix}
\etocsettagdepth{mtchapter}{none}
\etocsettagdepth{mtappendix}{subsection}
\etoctocstyle{1}{Contents (Appendix)}
\tableofcontents

\section{Derivation of the influence formula}
\label{si-sec:infl-derivation}

In this section, we detail the calculation of the tweet influence $\varphi(v)$, proposed in Sec.~\ref{sec:user-influence}.
In Sec.~\ref{subsec:diffusion-scenario}, we define the notion of diffusion scenario, and we compute its likelihood given an observed retweet cascade.
In Sec.~\ref{subsec:user-influence}, we compute the formula for tweet influence over all possible diffusion scenarios associated with the given cascade.

\subsection{Diffusion scenarios}
\label{subsec:diffusion-scenario}

\begin{figure*}[tbp]
	\newcommand\mywidth{0.5}
	\newcommand\myheight{0.2}
	\centering
	
	\subfloat[] {
		\begin{tabular}{c}
			\includegraphics[width=0.32\textwidth,valign=c]{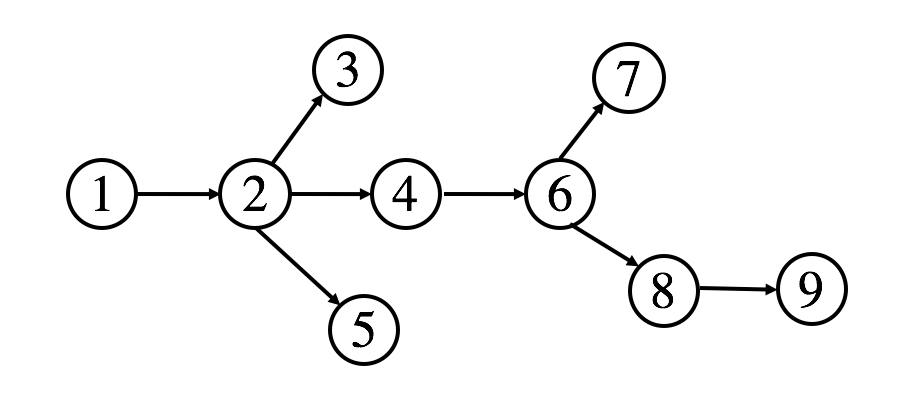} \\
			\includegraphics[width=0.51\textwidth,valign=c]{onediff}
		\end{tabular}
		\label{fig:example-diffusion-graph-construction}
	}
	\subfloat[] {
			\includegraphics[height=0.2\textheight,valign=c]{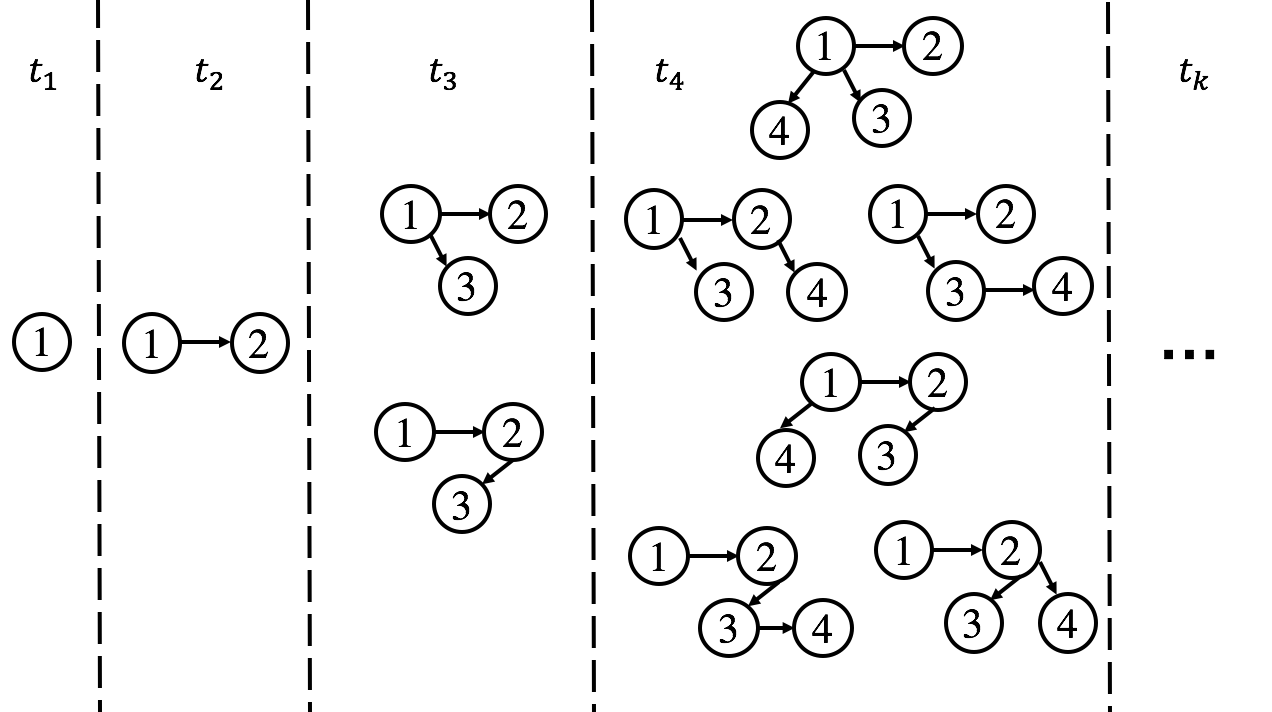}
			\vphantom{ 
				\begin{tabular}{c}
					\includegraphics[width=0.32\textwidth,valign=c]{diffgraph} \\
					\includegraphics[width=0.51\textwidth,valign=c]{onediff}
				\end{tabular}
			}
			\label{fig:diffusion-scenarios} 
		}	
	
%

	\caption{
		\textbf{(a)} An example of a diffusion tree \textbf{(top)} and its corresponding incremental diffusion process \textbf{(bottom)}.
		\textbf{(b)} Enumeration of all possible diffusion scenarios: 
		each retweet $v_k$ arrives at time $t_k$, and it can attach to any of the previous nodes, in any of the diffusion scenario constructed at time $t_{k-1}$.
		At time $t_k$ there are $(k-1)!$ diffusion scenarios, each with $k$ nodes.
	}
\end{figure*}

\textbf{Diffusion trees.}
We can represent an online diffusion using a directed tree $G(V, E)$, in which each node has a single parent and the direction of the edges indicates the flow of the information.
For retweet cascades, the nodes $v \in V$ are individual tweets and each directed edge $e \in E, e = \{v_a, v_b\}$ (showing the direction $v_a \longrightarrow v_b$) indicates that $v_b$ is a \emph{direct retweet} of $v_a$.
A direct retweet means that $u_b$ -- the user that emitted the tweet $v_b$ -- clicked on the ``Retweet'' option under tweet $v_a$.
The top panel of Fig.~\ref{fig:example-diffusion-graph-construction} shows an example of such a diffusion tree.
Note that each node $v$ has associated a time of arrival $t_v$ and that the diffusion tree respects the order of the times of arrival -- i.e. given the edge $e = \{v_a, v_b\}$, then $t_a < t_b$.
The bottom panel of Fig.~\ref{fig:example-diffusion-graph-construction} shows the incremental construction of the diffusion tree shown in top panel:
node $v_1$ is the root of the tree and the source of the information diffusion; 
at each time $t_i$, node $v_i$ attaches to the previous tree constructed at time $t_{i-1}$.



\begin{table}[!b]
\caption{Summary of notations.}
\small
\centering
\begin{tabular}{cp{5.5cm}}
\toprule
Notation & Interpretation \\ 
\midrule
	$G(V, E)$ & diffusion tree. In case the tree is unobserved, $G$ is a diffusion scenario. \\ 
	$v \in V$ & node in the diffusion tree (i.e. retweets). \\ 
	$e = \{v_a, v_b\} \in E$ & directed edge in the diffusion tree, tweet $v_b$ is a direct retweet of $v_a$.\\
	$t_v$ & time of arrival of node $v$ (timestamp of the tweet). \\
	$u_v$ & user that has emitted tweet $v$. \\
	$m_v$ & local influence (i.e. number of followers) of user $u_v$. \\
\bottomrule
\end{tabular}
\label{tab:parameters}
\end{table}

\textbf{Diffusion scenarios for retweet cascades.}
The diffusion tree is not observed for real Twitter retweet cascades, since the Twitter API does not expose the direct retweet relationships.
Instead, it assigns every retweet in the cascade to the original tweet.
%
Every retweet cascades constructed based on raw retweet information from the Twitter API resembles the graph in Fig.~\ref{fig:side:a}.
Due to this particular shape, we denote retweet cascades as \emph{stars}.
However, the API exposes the time of arrival of the retweets $t_i$.
We denote as a \emph{diffusion scenario} any valid diffusion tree that could be associated with the observed retweet star -- i.e., the edges in the diffusion tree respects the order of arrival of retweets.
%
Fig.~\ref{fig:side:b} shows four examples of diffusion scenarios associated with the star in Fig.~\ref{fig:side:a}.


\textbf{Constructing diffusion scenarios.}
Fig.~\ref{fig:diffusion-scenarios} exemplifies a straight-forward method to enumerate all diffusion scenarios associated with the star in Fig.~\ref{fig:side:a}.
The node $v_1$ is the root node and it is published at time $t_1$;
tweet $v_2$ occurs at time $t_2$ and it is undoubtedly a direct retweet of $v_1$ -- a directed edge is drawn from $v_1$ to $v_2$. 
Tweet $v_3$ observed at $t_3$ can be a direct retweet of either tweet $v_1$ or tweet $v_2$.
Therefore, at time $t_3$ there are two possible diffusion scenarios: 
$G_1$ with the edge set $E = \lbrace \{v_1, v_2\}, \{v_1, v_3\} \rbrace$ and $G_2$ with the edge set $E = \lbrace \{v_1, v_2\}, \{v_2, v_3\} \rbrace$.
Similarly, $v_4$ can be a direct retweet of $v_1$, $v_2$ or $v_3$, in either $G_1$ or $G_2$.
Consequently, at time $t_4$ there are 6 possible diffusion scenarios.
The process continues until all nodes have been attached.
For an observed star of size $k$, there are $(k-1)!$ associated diffusion scenarios.

\textbf{Probability of a diffusion scenario.}
Given that in retweet cascades individual edges are not observed,
we define the probability of an edge $\mathds{P}(\{v_a, v_b\})$ as the likelihood that $u_b$ emitted tweet $v_b$ as a direct retweet of $v_a$.
We model two factors in the likelihood of retweeting:
%
%
firstly, users retweet \emph{fresh} content~\cite{Wu2007}.
The probability of the edge $e$ decays exponentially with the time difference $t_b - t_a$;
secondly, users prefer to retweet locally influential users, also known as preferential attachment~\cite{Barabasi2005}.
We measure the local influence of a user using his number of followers~\cite{kwak2010twitter,Cha2010,Rizoiu2017}.
We quantify the probability of an edge as:
\begin{equation} \label{eq:prob-edge}
	\mathds{P}(\{v_a, v_b\}) = m_a e^{-r({t_{b}-t_{a}})}
\end{equation}
where 
$m_a$ is the number of followers of the user of $u_a$ and
$r$ controls the temporal decay of the probability.

Under the assumption that retweeting events (i.e. edges) occur independently one from another, we obtain the probability of a diffusion scenario as:
%
\begin{equation} \label{eq:prob-scenario}
	\mathds{P}(G) = \prod_{\{v_a,v_b\} \in E} \mathds{P}(\{v_a, v_b\})
\end{equation}

Note that the above assumption of independence of retweet events is a strong assumption.
Current state-of-the-art approaches~\cite{Mishra2016,Zhao2015} for modeling retweet cascades employ self-exciting point processes, in which the arrival of one event increases the probability of future events.
However, for our application of estimating the probability of a diffusion scenario it is the simplest assumption.
Additional arguments in its favor are also that we are studying networks of events
(in which each event is identified with unique user), not networks of users. 
The interdependence often observed in socially generated networks (like triadic closure)
can be ignored in edge formation within a particular retweet cascade. 

\subsection{Computing tweet influence}
\label{subsec:user-influence}

\cite{Du2013} define user influence of a user $u$ as the average number of users in the social network who get in contact with the content emitted by $u$.
For retweet cascades the diffusion tree is not observed; it is impossible to directly measure user influence, apart from the root user.
We define the \emph{tweet influence} over a retweet cascade as the expected number of time it is retweeted -- direct retweets or descendants in a diffusion scenario --, over all possible diffusion scenarios associated with the given star.
Finally, we compute the influence of user $u$ as the sum of the influences of the tweets that $u$ authored.
We see that the definition in~\cite{Du2013} is a special case of our definition, in which the diffusion tree is observed.

\textbf{Tweet influence over one diffusion scenario.}
%
Let $z(v_i, v_k) \in G$ be a path in the diffusion scenario $G$ -- i.e a sequence of nodes which starts with $v_i$ and ends with $v_k$.
$\mathds{P}(z(v_i, v_k))$ is the probability of reaching $v_k$ from $v_i$.
$\varphi(v_i | G)$ is the influence of $v_i$ and it is computed as the expected number of users reached from $v_i$ using a model of independent binomials to decide whether or not to take each hop in each path that starts with $v_i$.
Formally:
%
\begin{align}
	\varphi(v_i|G) 	&= \mathds{E} \left[ \sum_{v_k \in V(G)} \mathds{1} \left\lbrace z(v_i,v_k|G) \right\rbrace \right] \nonumber \\
					&= \sum_{v_k \in V(G)} \mathds{E}\bigg[\mathds{1} \left\lbrace z(v_i,v_k|G) \right\rbrace \bigg] \nonumber \\
					&= \sum_{v_k \in V(G)} \mathds{P}(z(v_i,v_k|G)) \label{eq:infl-scenario-path} \\
					&= \sum_{v_k \in V(G)} \prod_{\{v_a,v_b\} \in z(v_i, v_k)} \mathds{P}\big(\{v_a, v_b\}\big) \label{eq:infl-in-a-scenario}
\end{align}
where $\mathds{1} \left\lbrace z(v_i,v_k|G) \right\rbrace$ is a function that takes the value 1 when the path from $v_i$ to $v_k$ exists in $G$.

\textbf{Tweet influence over a retweet cascade.}
We compute the influence of tweet $v_i$ over $VG$ -- all possible diffusion scenarios associated with a retweet cascade -- as:
\begin{align}
	\varphi(v_i) 	&= \sum_{G\in{VG}} \mathds{P}(G) \varphi(v_i|G) \nonumber \\
				&= \sum_{G\in{VG}} \mathds{P}(G) \sum_{v_k \in V(G)} \mathds{P}\big( z(v_i,v_k|G) \big) \label{eq:brute-user-infl}
\end{align}
It is intractable to directly evaluate Eq.~\eqref{eq:brute-user-infl}, plugged in with Eq.~\eqref{eq:prob-scenario} and~\eqref{eq:infl-in-a-scenario}, particularly due to the factorial number of diffusion scenarios in $VG$.
For example, there are $10^{156}$ diffusion scenarios for a cascade of 100 retweet.
We develop, in the next section, an efficient linear time algorithm to compute the influence of all tweets in a retweet cascade.

\section{Efficient tweet influence computation}
\label{si-sec:efficient-algo}

The key observation is that each tweet $v_k$ is added simultaneously at time $t_k$ to all diffusion scenarios constructed at time $t_{k-1}$.
$v_k$ contributes only once to the tweet influence of each tweet that is found on the branch it attached to.
This process is exemplified in Fig.~\ref{fig:add-one-edge}.
Node $v_5$ is added to a given diffusion scenario, generating 4 new diffusion scenarios at time $t_5$.
We color in red the nodes whose influence increases as a result of adding node $v_5$.
%
This allows to compute the tweet influence incrementally, by updating $\varphi(v_i), i < k$ at each time $t_k$.
We denote by $\varphi^k(v_i)$ the value of tweet influence of $v_i$ after adding node $v_k$.
As a result, we only keep track of how tweet influence increases over time steps and we do not require to construct all diffusion scenarios.

\subsection{Complete derivation of the recursive influence formula}

\textbf{Incremental construction of diffusion scenarios.}
Let $G^- \in VG^{k-1}$ be a diffusion scenario constructed at time $t_{k-1}$, with the set of nodes $V^- = \{ v_1,v_2,\cdots,v_{k-1} \}$.
When $v_k$ arrives, it can attach to any node in $V^-$, generating $k-1$ new diffusion scenarios $G_j^+$, with $V_j^+ = V^- \cup v_k$ and $E_j^+ = E^- \cup \{v_j, v_k\}$.
This process is exemplified in Fig.~\ref{fig:add-one-edge}.
We can write the set of scenarios at time $t_k$ as:
\begin{equation} \label{eq:diff-scenario-increase}
	VG^k = \left\lbrace G_j^+ = G^- \cup \{v_j, v_k\} \middle| \forall j < k, \forall G^- \in VG^{k-1} \right\rbrace
\end{equation}     
We write the tweet influence of $v_i$ at time $k$ as:
%
%
%
\begin{align}
	\varphi^k(v_i) &= \sum_{G^+ \in {VG^k}} \mathds{P}(G^+) \varphi(v_i|G^+) \nonumber \\
	^{cf.~\eqref{eq:diff-scenario-increase}} &= \sum_{G^- \in {VG^{k-1}}}\sum^{k-1}_{j=1}\mathds{P}(G_j^+) \varphi(v_i|G_j^+) \label{eq:infl-step-k}
\end{align}
Note that in Eq.~\eqref{eq:infl-step-k} we explicitly make use of how diffusion scenarios at time $k$ are constructed based on the diffusion scenarios at time $k-1$.

\textbf{Attach a new node $v_k$.}
We concentrate on the right-most factor in Eq.~\eqref{eq:infl-step-k} -- the tweet influence in scenario $G^+_j$.
We observe that the terms in Eq.~\eqref{eq:infl-scenario-path} can be divided into two:
the paths from $v_i$ to all other nodes except $v_k$ and the path from $z(v_i, v_k)$.
We obtain:
%
%
\begin{equation*}
	\varphi(v_i|G_j^+) = \sum_{\substack{v_l\in{G_j^+}\\l>i, l\neq k}} \mathds{P}(z(v_i,v_l|G_j^+)) + \mathds{P}(z(v_i,v_k|G_j^+)).
\end{equation*}
Note that a path that does not involve $v_k$ has the same probability in $G_j^+$ and in its parent scenario $G^-$:
\begin{equation*}
	\mathds{P}(z(v_i,v_l|G_j^+)) = \mathds{P}(z(v_i,v_l|G^-)), \text{ for } l > i, l \neq k
\end{equation*}
we obtain:
\begin{align}
				\varphi(v_i|G_j^+) 		&= \sum_{\substack{v_l\in{G^-}\\l>i}} \mathds{P}(z(v_i,v_l|G^-)) + \mathds{P}(z(v_i,v_k|G_j^+)) \nonumber \\
^{cf.~\eqref{eq:infl-scenario-path}}	&= \varphi(v_i | G^-) + \mathds{P}(z(v_i,v_k|G_j^+)) \label{eq:infl-step-k+1}
\end{align}
Combining Eq.~\eqref{eq:infl-step-k} and~\eqref{eq:infl-step-k+1}, we obtain:
\begin{align}
\varphi^k(v_i) &= \sum_{\substack{G^-\\\in VG^{k-1}}}\sum^{k-1}_{j=1} \mathds{P}(G_j^+) \bigg[ \varphi(v_i | G^-) + \mathds{P}(z(v_i,v_k|G_j^+)) \bigg] \nonumber \\
        &= \underbrace{\sum_{G^- \in {VG^{k-1}}} \varphi(v_i | G^-) \sum^{k-1}_{j=1}\mathds{P}(G_j^+) }_{A} \nonumber \\
        &+ \underbrace{\sum_{G^- \in {VG^{k-1}}}\sum^{k-1}_{j=1} \mathds{P}(G_j^+) \mathds{P}(z(v_i,v_k|G_j^+)) }_{M_{ik}}. \label{eq:two-parts}
\end{align}

\textbf{Tweet influence at previous time step $t_{k-1}$.}
Given the definition of $G_j^+$ in Eq.~\eqref{eq:diff-scenario-increase}, we obtain that $\mathds{P}(G_j^+) = \mathds{P}(G^-)\mathds{P}(\{v_j,v_k\})$.
Consequently, part $A$ in Eq.~\eqref{eq:two-parts} can be written as:
\begin{align}
	A 	&= \sum_{G^- \in {VG^{k-1}}} \varphi(v_i | G^-) \sum^{k-1}_{j=1} \mathds{P}(G^-)\mathds{P}(\{v_j,v_k\}) \nonumber \\
		&= \sum_{G^- \in {VG^{k-1}}} \varphi(v_i | G^-) \mathds{P}(G^-) \sum^{k-1}_{j=1} \mathds{P}(\{v_j,v_k\}) \nonumber \\
		&= \sum_{G^- \in {VG^{k-1}}} \mathds{P}(G^-) \varphi(v_i | G^-)  \nonumber \\
	^{cf.~\eqref{eq:brute-user-infl}}	&= \varphi^{k-1}(v_i) \label{eq:past-infl}
\end{align}
Consequently, $A$ is the tweet influence of $v_i$ at the previous time step $t_{k-1}$.
Note that $\sum^{k-1}_{j=1} \mathds{P}(\{v_j,v_k\}) = 1$ because $v_k$ is necessarily the direct retweet of one of the previous nodes $v_j, j < k$ of the retweet cascade.
%


\begin{figure*}[tbp]
	\centering
	\begin{minipage}{\textwidth}
		 \[
 \underbrace{\left[
\begin{matrix}
 \textcolor{gray}{1}  &\textcolor{blue}{M_{12}}  &\textcolor{red}{M_{13}}   &\textcolor{ForestGreen}{M_{14}}      & \cdots    & M_{1n} \\
 0                    &\textcolor{blue}{1}   	 &\textcolor{red}{M_{23}}   &\textcolor{ForestGreen}{M_{24}}      & \cdots    & M_{2n} \\
 0                    &0             			 &\textcolor{red}{1}   	    &\textcolor{ForestGreen}{M_{34}}      & \cdots    & M_{3n} \\
 0                    &0             			 &0             			&\textcolor{ForestGreen}{1}   		& \cdots    & M_{4n} \\
 \vdots    		      &\vdots            		 & \vdots       			&\vdots       				    & \ddots    & \vdots \\
 0         		      &0             			 &0                         &0             				    & \cdots    & 1      \\
\end{matrix}   
\right]}_{\textbf{M}} 
 \underbrace{\left[
\begin{matrix}
 1         &\textcolor{gray}{P_{12}}       &\textcolor{blue}{P_{13}}       &\textcolor{red}{P_{14}}        & \cdots    & P_{1n} \\
 0         &1             	    		   &\textcolor{blue}{P_{23}}       &\textcolor{red}{P_{24}}        & \cdots    & P_{2n} \\
 0         &0             				   &1					   		   &\textcolor{red}{P_{34}}         & \cdots    & P_{3n} \\
 0         &0             				   &0             				   &1					  		    & \cdots    & P_{4n} \\
 \vdots    &\vdots            			   & \vdots       				   & \vdots       				    & \ddots    & \vdots \\
 0         &0             				   &0                              &0             					& \cdots    & 1      \\
\end{matrix}   
\right]}_{\textbf{P}}
\]
\[
\begin{alignedat}{3}
&\begin{array}{rl}&M_{11} = 1 \end{array} &k =1\ \ \ \ \ &\\ 
&\begin{array}{rl}&M_{12} =P_{12}M_{11}\end{array} &k= 2 \ \ \ \ &\left[\textcolor{gray}{P_{12}}\right]\left[\textcolor{gray}{M_{11}}\right] = 
\left[\textcolor{blue}{M_{12}}\right]\\
&\begin{array}{rl}
	&M_{13} =P_{13}M_{11}+ P_{23}M_{12}\\
	&M_{23} =P_{13}M_{21}+ P_{23}M_{22}
\end{array}
\Bigg\} &k = 3 \ \ \ \ 
&\left[
\begin{matrix}
\textcolor{gray}{1}  &\textcolor{blue}{M_{12}} \\
0                    &\textcolor{blue}{1}   	
\end{matrix}
\right]
\left[
\begin{matrix}
\textcolor{blue}{P_{13}}\\
\textcolor{blue}{P_{23}}
\end{matrix}
\right]=
\left[
\begin{matrix}
\textcolor{red}{M_{13}}\\
\textcolor{red}{M_{23}}
\end{matrix}
\right]\\
&\begin{array}{rl}
	&M_{14} =P_{14}M_{11}+ P_{24}M_{12}+ P_{34}M_{13}\\
	&M_{24} =P_{14}M_{21}+ P_{24}M_{22}+ P_{34}M_{23}\\
    &M_{34} =P_{14}M_{31}+ P_{24}M_{32}+ P_{34}M_{33}
\end{array}
\Bigg\} &k = 4 \ \ \ \ 
&\left[
\begin{matrix}
 \textcolor{gray}{1}  &\textcolor{blue}{M_{12}}  &\textcolor{red}{M_{13}}\\
 0                    &\textcolor{blue}{1}   	 &\textcolor{red}{M_{23}}\\
 0                    &0             			 &\textcolor{red}{1}   	 
\end{matrix}
\right]
\left[
\begin{matrix}
\textcolor{red}{P_{14}}\\
\textcolor{red}{P_{24}}\\
\textcolor{red}{P_{34}}
\end{matrix}
\right]=
\left[
\begin{matrix}
\textcolor{ForestGreen}{M_{14}}\\
\textcolor{ForestGreen}{M_{24}}\\
\textcolor{ForestGreen}{M_{34}}
\end{matrix}
\right]\\
\end{alignedat}
\]
	\end{minipage}
	\caption{
		Exemplification of the efficient computation of the first four columns of matrix $M$.
		Each $k^{th}$ column vector of matrix $M$ is colored correspondingly with the column vector in $P$ used in the multiplication.
%
%
	}
	\label{fig:matrix-operations-example}
\end{figure*}

\textbf{Contribution of $v_k$.}
With $A$ being the influence of $v_i$ at the previous time step, intuitively $M_{ik}$ is the contribution of $v_k$ to the influence of $v_i$.
Knowing that:
\begin{align*}
	\mathds{P}(G_j^+) &= \mathds{P}(G^-)\mathds{P}(\{v_j,v_k\}) \text{ and } \\
	\mathds{P}\{z(v_i,v_k|G_j^+)\} &= \mathds{P}(z(v_i,v_j|G^+_j)) \mathds{P}(\{v_j,v_k\}) \\
	 &= \mathds{P}(z(v_i,v_j|G^-)) \mathds{P}(\{v_j,v_k\})
\end{align*}
we write $M_{ik}$ as:
%
\begin{align}
	M_{ik} &= \sum_{G^- \in {VG^{k-1}}} \sum^{k-1}_{j=1} \mathds{P}(G^-)\mathds{P}^2(\{v_j,v_k\}) \mathds{P}(z(v_i,v_j|G^-)) \nonumber \\
           &= \sum^{k-1}_{j=1} \mathds{P}^2(\{v_j,v_k\}) \underbrace{ \sum_{G^- \in {VG^{k-1}}} \mathds{P}(G^-) \mathds{P}(z(v_i,v_j|G^-))}_{M_{ij}} \nonumber \\  
		   &= \sum^{k-1}_{j=1} \mathds{P}^2(\{v_j,v_k\}) M_{ij} \label{eq:M-ik}
\end{align}
An alternative interpretation for $M_{ik}$ is the influence of $v_i$ over $v_k$.
Eq.~\eqref{eq:M-ik} can be intuitively understood as the expected influence of $v_i$ over a newly attached node $v_k$ is proportional to the influence of $v_i$ over each already attached node $v_j$ multiplied with the likelihood that $v_k$ attached itself to $v_j$.
We obtain the formula of the the expected influence of a node $v_i$ over another node $v_k$ as:
%
\begin{equation} \label{eq:Mij}
M_{ik}=
\left\{
\begin{array}{ll}
	\sum^{k-1}_{j=1}M_{ij}\mathds{P}^2(\{v_j, v_k\}) &,i < k \\
	1 & ,i = k \\
	0 & ,i > k.
\end{array}
\right.
\end{equation} 

\textbf{Recursive computation of tweet influence.}
Using Eq~\eqref{eq:two-parts} and~\eqref{eq:past-infl}, we can recursively compute the influence of $v_i$ at time $t_k$ as:
\begin{align}
	\varphi^k(v_i) &= \varphi^{k-1}(v_i) + M_{ik} \nonumber \\
				   &= \varphi^{k-2}(v_i) + M_{i(k-1)} + M_{ik} \nonumber \\
				   &= \sum_{j=1}^k M_{ij}. 
\end{align}
with $M_{ij}$ as defined in Eq.~\eqref{eq:Mij}.
Intuitively, the influence of $v_i$ at time $t_k$ is the sum of the expected influence over each of the nodes in the diffusion scenario.


\subsection{Tweet influence algorithm}

We define two matrices:
matrix $P = [ P_{ij} ]$, with $P_{ij} = \mathds{P}^2(\{v_i, v_j\})$.
Element $P_{ij}$ of the matrix $P$ is the square probability that tweet $v_j$ is a direct retweet of tweet $v_i$;
matrix $M = [ M_{ij} ]$, with $M_{ij}$ defined in Eq.~\eqref{eq:Mij}.
Element $M_{ij}$ of matrix $M$ is the contribution of $v_j$ to the influence of $v_i$.
Alternatively, $M_{ij}$ can be interpreted as the influence of $v_i$ on $v_j$.

From Eq.~\eqref{eq:Mij} follows the formula for computing iteratively the columns of matrix $M$:
\begin{equation} \label{eq:Mij-matrix}
M_{ \cdot j}=
\left[
\begin{array}{c}
M_{[1..j-1, 1..j-1]} \times P_{[1:j-1,j]} \\
1 \\
0 \\
0 \\
\vdots \\
0 
\end{array}
\right]
\end{equation}
with the value 1 occurring on line $j$.
For each column $j$, we compute the first $j-1$ elements by multiplying the sub-matrix $M_{[1..j-1, 1..j-1]}$ with the first $j-1$ elements on the $j^{th}$ column of matrix $P$.
The computation of matrix $M$ finishes in linear time, after $n$ steps, where $n$ is the total number of retweets in the retweet cascade.
Fig.~\ref{fig:matrix-operations-example} demonstrated the computation of the first four columns in $M$.
Algorithm~\ref{alg:casin} gives an overview of the efficient influence computation algorithm.

\begin{figure*}[tbp]
	\centering
	\newcommand\mywidth{0.3}
	\subfloat[] {
		\includegraphics[width=\mywidth\textwidth, page=1]{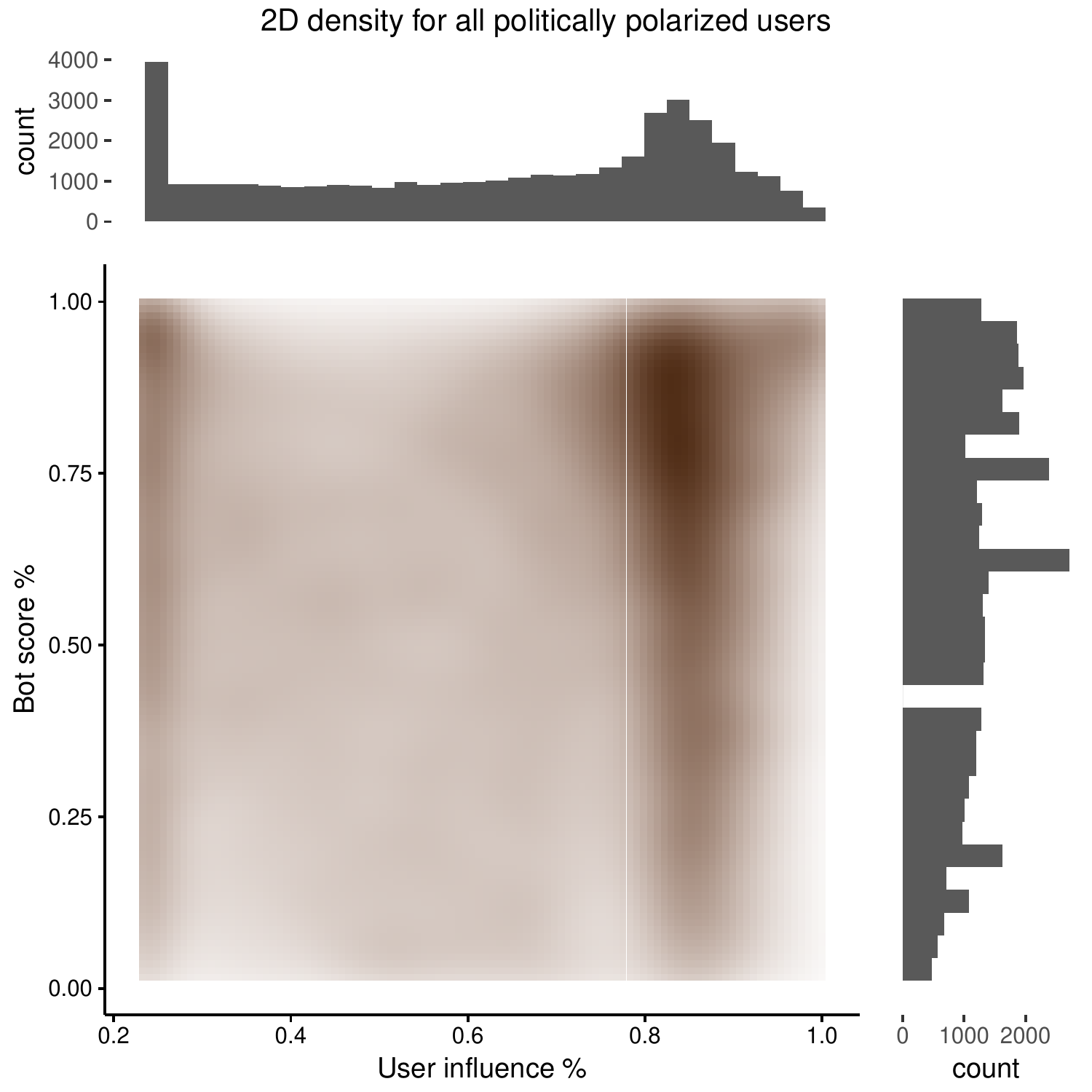}
	}
	\subfloat[] {
		\includegraphics[width=\mywidth\textwidth, page=2]{2017-11-28-2d-density-map-influence-botscore}
	}
	\subfloat[] {
		\includegraphics[width=\mywidth\textwidth, page=3]{2017-11-28-2d-density-map-influence-botscore}
	}
	\caption{
		2D density plots for all politically polarized users in \debate \textbf{(a)}, for Democrat users \textbf{(b)} and for Republican users \textbf{(c)}.
	}
	\label{si-fig:additional-2d}
\end{figure*}

\section{Generation of synthetic data}
\label{si-sec:generation-artificial}

This section completes and details the results concerning the evaluation on synthetic data, presented in the main text Sec.~\ref{subsec:ground-truth}.
This section details the construction of a synthetic random social graph (Sec.~\ref{si-subsec:random-graph}) and the sampling of synthetic cascades (Sec.~\ref{si-subsec:random-sampling}).
The purpose is to construct a synthetic dataset of cascades, in which the user influence ground truth is known.
Both the graph and cascade generators described here below reproduce closely the synthetic experimental setup described in~\cite{Du2013}.

\subsection{Generation of random graphs}
\label{si-subsec:random-graph}

In this section, we describe the construction of a synthetic social graph, with $n$ nodes, specified by its adjacency matrix $M$.
Each node corresponds to a synthetic user, and the edges correspond to synthetic follow relations.
We follow the below steps:
\begin{itemize}
	\item Given the number of nodes $n$ in the graph (here $1000$), create an null (all zero) adjacency matrix $M$ of size $n \times n$;
	\item Randomly choose the number of edges $|E|$, between $n/2$ to $n^2$.
	\item Randomly choose $i$ and $j$ between $1$ to $n$ and set $M_{ij}$ to $1$. Iterate this step until $|E|$ different edges are generated.
	\item The adjacency matrix $M$ defines the final random graph.
\end{itemize}

\subsection{Sampling synthetic cascades}
\label{si-subsec:random-sampling}

In this section we describe how to construct synthetic cascades, given a synthetic social graph $G$ constructed as shown in Sec.~\ref{si-subsec:random-graph}.
To generate one cascades, to each edge in $G$ we associate an exponentially distributed waiting time and we construct a shortest path tree.
We detail this procedure in the following steps:

\begin{itemize}
	\item Similar to previous work (ConTinEst~\cite{Du2013}), for each edge $\{v_i, v_j\} \in G$ draw a transmission rate $r_{ij}$ from a Weibull distribution of shape $k = 2$.
	\item Given the transmission rate $r_{ij}$, we draw an Exponentially distributed waiting time $\tau_{ij}$ using the inverse transform sampling:
		$$\tau_{ij} = -\frac{ln(s)}{r_{ij}}$$  
	where $s$ is draw uniformly from $U(0,1)$.
	\item Set $\tau_{ij}$ as the weight of edge $\{v_i, v_j\}$.
	\item Starting from a source node $v_s$, construct the shortest path tree from $v_s$ to all the other nodes in $G$;
	\item For each node $v_k$ compute two measures:
		$t_k$ -- its time of occurrence as the total waiting time along the path from $v_s$ to $v_k$ -- and
		$c_k$ -- the number of reachable nodes from $v_k$ in the shortest path tree;
	\item The generated cascade is $\{t_1,t_2,t_3 \cdots t_n\}$  where $(t_1 < t_2 < t_3 < \cdots t_n)$;
	\item The ground truth influence of node $v_k$ is the mean $c_k$ over multiple random graphs (here 100).
\end{itemize}

\section{Choosing the temporal decay parameter}
\label{si-sec:choose-temp-decay}

The temporal decay parameter $r$ shown in Eq.~\eqref{eq:prob-edge-mt} is determined by linear search. 
Eq.~\eqref{eq:prob-edge-mt} measures the probabilities of edges in the retweeting (diffusion) network, however the real diffusions are not observed.
We make the assumption that diffusions occur along edges in the underlying social graph (follower relation).
We measure the fitness of edge probability by the likelihood of uncovering the ground truth follower graph.
In other words, if the edge $\{v_i, v_j\}$ exists in the follower graph (i.e. $v_j$ is a follower of $v_i$) then the edge $\{v_i, v_j\}$  has the highest probability in the diffusion tree than the edge $\{v_l, v_j\}$ which does not exist in the following graph.
In other words, we are using the retweeting probability to predict the existence of edges in the social graph.
We randomly select 20 cascades, and we crawl the following list of every user appearing the the diffusions (the following list for a user $u_j$ consists of users $u_i$ followed by $u_j$).
We use this information as ground truth for the following prediction exercise:
given a user $u_j$ who emitted tweet $v_j$ in a particular cascade, we want to predict which among the users in the set $\{u_1, u_2, \ldots, u_{j-1}\}$ are followed by $u_j$.
Considering that $\mathds{P}(\{v_l, v_j\}). \forall l < j, j \geq 2$ are real numbers and the prediction target is binary, we use the AUC (are under ROC curve) the measure the prediction performance of a particular probability scoring function.
For each value of $r$ in Eq.~\eqref{eq:prob-edge-mt} we compute the mean AUC over all predictions.
We perform a linear search for the optimal $r$ between $10^{-8}$ to $3$. 
Finally, $6.2\times10^{-4}$ maximizes the mean of AUC and it is chosen as $r$ value in the experiments in Sec.~\ref{sec:evaluation-influence} and~\ref{sec:results-findings}.
%
%
%

\section{Additional 2D densities plots}
\label{si-sec:polarization-map}

We show in Fig.~\ref{si-fig:additional-2d} the additional 2D density plots mentioned in the main text Sec.~\ref{subsec:user-influence-results}.

\end{document}